\documentstyle[aps,multicol,eqsecnum]{revtex}

\begin{document}
\title{Exact Diagonalization of Two Quantum Models for the Damped Harmonic
Oscillator}
\author{M.\ Rosenau da Costa\thanks{E-mail: rosenau@ifi.unicamp.br}, A.\ O.\ Caldeira\thanks{E-mail: caldeira@ifi.unicamp.br}, 
S.\ M.\ Dutra\thanks{Present address: Huygens Laboratory, University of Leiden, 
P.\ O.\ Box 9504, 2300 RA Leiden, The Netherlands. 
E-mail: dutra@molphys.leidenuniv.nl}, H.\ Westfahl, Jr.
\thanks{Present address: Department of Physics of Illinois at Urbana-Champaign,
 1110 West Green Street, Urbana, IL 61801-3080, USA. 
E-mail: westfahl@cromwell.physics.uiuc.edu}\\
\emph{Instituto de F\'{\i}sica ``Gleb Wataghin''}\\
\emph{Universidade Estadual de Campinas, Unicamp} \\
\emph{Caixa Postal 6165, 13083-970 Campinas, S\~{a}o Paulo, Brazil}}
\date{\today}
\maketitle

\begin{abstract}
The damped harmonic oscillator is a workhorse for the study of dissipation in 
quantum mechanics. However, despite its simplicity, this system has given rise 
to some approximations whose validity and relation to more refined descriptions 
deserve a thorough investigation. In this work, we apply a method that allows 
us to diagonalize exactly the dissipative Hamiltonians that are frequently 
adopted in the literature. Using this
method we derive the conditions of validity of the
rotating-wave approximation (RWA) and show how this approximate description relates to
more general ones. We also show that the existence of
dissipative coherent states is intimately related to the RWA. Finally,
through the evaluation of the dynamics of the damped oscillator, we notice 
an important property of the dissipative model that has not been
properly accounted for in previous works; namely, the necessity of new
constraints to the application of the factorizable initial conditions.
\end{abstract}

\section{Introduction}
The study of dissipative systems and in particular of the Brownian motion
has been pursued for a long time in the context of classical \cite{clas} and
quantum mechanics \cite{AmB}. Although there has been a number of
publications in this area there are some subtle points that have never been properly investigated 
in the literature. Among these we could
mention three major ones; a careful investigation of the relation
between different models \cite{Ford}, the existence of
dissipative coherent states \cite{Agar,Wl1,Wl2,Ser} or the
condition for the employment of factorizable initial conditions. These are
exactly the issues we shall address in this paper.

Usually the dissipation in the system is described as a consequence of its
coupling to a reservoir. The properties of this dissipative systems are
generally studied through the evaluation of the time evolution of its
reduced density operator. This evolution is often described either by a
generalization of the Feynman-Vernon approach \cite{AmB,AmA,Grab,Cris} or through master
equations \cite{Agar,Wl1,Wl2,Ser,Pab,Zur,Zur0,Pab2,Zur2}. In this work the
properties of the system will be studied through exact diagonalization of
different Hamiltonians of the dissipative models.

We will consider a damped harmonic oscillator. The usual models of
dissipation consist of coupling the harmonic oscillator to a reservoir that
is conveniently chosen as a group of N noninteracting oscillators. The
coupling between the two systems is bilinear in the creation and destruction
operators of quanta of energy. Then the Hamiltonian of the total system is
given by \cite{Agar} 
\begin{equation}
\hat{H}=\hat{H}_{Sis}+\hat{H}_{Res}+\hat{H}_{Int},  \label{0}
\end{equation}
being 
\begin{eqnarray}
\hat{H}_{Sis} &=&\hbar \omega _{o}\hat{a}^{\dagger }\hat{a},\quad \hat{H}%
_{Res}=\hbar \sum_{j}\omega _{j}\hat{b}_{j}^{\dagger }\hat{b}_{j},  \label{1}
\\
\hat{H}_{Int} &=&\hbar \left( \hat{a}^{\dagger }+\hat{a}\right)
\sum_{j}\left( k_{j}\hat{b}_{j}+k_{j}^{*}\hat{b}_{j}^{\dagger }\right) ,
\label{2}
\end{eqnarray}
where we consider a harmonic oscillator with frequency $\omega _{o}$ (the
system of interest) interacting with a bath of oscillators with frequencies $%
\omega _{j}$ through the coupling constants $k_{j}$'s. We will take the
limit of a continuous spectrum of excitations in the reservoir of the
Hamiltonian $\hat{H}$. Then we will diagonalize $\hat{H}$
and determine the time evolution of the operator $\hat{%
a}$ exactly. The analysis of $\hat{a}\left( t\right) $ will determine the
conditions of validity of the rotating-wave approximation (RWA) which
consists of neglecting the terms $k_{j}\hat{a}\hat{b}_{j}+k_{j}^{*}\hat{a}%
^{\dagger }\hat{b}_{j}^{\dagger }$ in (\ref{1}) and writing
\begin{equation}
\hat{H}_{Int}^{RWA}=\hbar \sum_{j}\left( k_{j}\hat{a}^{\dagger }\hat{b}%
_{j}+k_{j}^{*}\hat{a}\hat{b}_{j}^{\dagger }\right) .  \label{3}
\end{equation}

Once this has been accomplished we will discuss the existence
of dissipative coherent states. Some authors \cite{Agar,Wl1,Wl2,Ser} 
have stated that the coherent states are special states that
remain pure during its decay in dissipative systems. We will show
that the existence of these dissipative coherent states is directly related
to the RWA; they can only exist at zero temperature and in systems that meet the 
conditions required for the RWA.

Once we have determined the evolution of the operator $\hat{a}\left(
t\right) $ of the system we can determine the evolution of any of its
observables. However, the dynamics of these observables will depend on the
specific form adopted for the coupling constants $k_{j}$ as 
functions of the frequencies $\omega _{j}$. Our method holds for an arbitrary 
form but in order to compare our results with the Caldeira-Leggett 
model \cite{AmB}, we will reduce our results to the case where the function 
becomes the same as the one they have 
adopted. Then, as in Refs. \cite{Grab,Cris}, we will determine the evolution of the mean value
of the position operator $\left\langle \hat{q}\left( t\right) \right\rangle $
of the damped oscillator. The result of this calculation reveals a very special need to carefully
treat the initial time of the motion. We propose a simple initial condition
that eliminates the initial transient that would appear in the evolution of 
$\left\langle \hat{q}\left( t\right) \right\rangle $ and we believe that it is enough to eliminate
most of or maybe all the initial transients (in a certain time scale) which
were noticed in this system in previous works \cite{Pab,Zur,Pab2}.

The paper is organized as follows. In Sec. II we write the Hamiltonian (\ref
{0}) in the limit of a continuous spectrum for the reservoir excitations and
we diagonalize it exactly within and without the RWA. We compare the model
given by the Hamiltonian (\ref{0}) with the dissipative model presented in 
\cite{AmB} in Sec. III. Here, we also determine the relation between the
coupling function $\left| v\left( \omega \right) \right| ^{2}$, introduced
in Sec. II, and the spectral function $J\left( \omega \right) $ introduced
in \cite{AmA}. In Sec. IV we analyze the relevance of the different terms
that appear in the calculation of the evolution of the operator $\hat{a}%
\left( t\right) $ with relation to the intensity of the dissipation in the
system. In Sec. V we show under which conditions the evolution of the
operator $\hat{a}\left( t\right) $ is reduced to that given in the RWA. In
Sec. VI we show that the existence of dissipative coherent states is only
possible within the RWA. In Sec. VII we present the calculation of the
evolution of the mean value of the position of the damped harmonic
oscillator. In Sec. VIII we discussed the physical meaning of the initial condition
proposed in Sec.VII. Finally, we discuss the main results and conclusions in
Sec. IX.

\section{Diagonalization of the Dissipative Hamiltonians}
\subsection{Treating a Reservoir with Continuous Spectrum}
We can rewrite the Hamiltonian (\ref{0}) considering a continuous spectrum
of excitations in the reservoir by making use of the transformation between
the discrete boson operators $\hat{b}_{j}$ and the continuous ones $\hat{b}%
_{\Omega }$ \cite{Coh}

\begin{equation}
\hat{b}_{j}=\sqrt{g\left( \Omega _{j}\right) }\int_{1/g\left( \Omega
_{j}\right) }d\Omega \hat{b}_{\Omega },  \label{230}
\end{equation}
where $g\left( \Omega _{j}\right) d\Omega _{j}$ is the number of modes in
the reservoir with frequencies between $\Omega _{j}$ and $\Omega
_{j}+d\Omega _{j}$ and $\int_{1/g\left( \Omega _{j}\right) }d\Omega $
represents an integration in a band of width $1/g\left( \Omega _{j}\right) $
around $\Omega _{j}$. The operators $\hat{b}_{\Omega }$ then satisfy the
commutation relation
\begin{equation}
\left[ \hat{b}_{\Omega },\hat{b}_{\tilde{\Omega}}^{\dagger }\right] =\delta
\left( \Omega -\tilde{\Omega}\right) ,  \label{230b}
\end{equation}
and all other commutators vanish.

Under the transformation (\ref{230}) we find 
\begin{equation}
\hat{H}_{Int}=\hbar \left( \hat{a}^{\dagger }+\hat{a}\right) \int d\Omega 
\sqrt{g\left( \Omega \right) }\left[ k\left( \Omega \right) \hat{b}_{\Omega
}+k^{*}\left( \Omega \right) \hat{b}_{\Omega }^{\dagger }\right] ,
\label{232}
\end{equation}
where we considered that $g\left( \Omega _{j}\right) $ and $k\left( \Omega
_{j}\right) $ are constant inside the interval $1/g\left( \Omega _{j}\right) 
$ and that $\sum_{j}\int_{1/g\left( \Omega _{j}\right) }d\Omega $ is nothing
but $\int d\Omega $, where this last integral covers the whole spectrum of
excitations of the reservoir. Then the total Hamiltonian of our system is given by 
\begin{eqnarray}
\hat{H} &=&\hbar \omega _{o}\hat{a}^{\dagger }\hat{a}+\hbar \int \Omega \hat{%
b}_{\Omega }^{\dagger }\hat{b}_{\Omega }d\Omega  \label{235} \\
&&\qquad \qquad +\hbar \left( \hat{a}^{\dagger }+\hat{a}\right) \int \left[
v\left( \Omega \right) \hat{b}_{\Omega }+v^{*}\left( \Omega \right) \hat{b}%
_{\Omega }^{\dagger }\right] d\Omega .  \nonumber
\end{eqnarray}
where 
\begin{equation}
v\left( \Omega \right) =\sqrt{g\left( \Omega \right) }k\left( \Omega \right)
.  \label{234}
\end{equation}

\subsection{The Hamiltonian within the Rotating Wave Approximation}
We will now perform a canonical transformation and apply the procedure proposed
by Fano \cite{Fn} in order to diagonalize the Hamiltonian of our global
system in the RWA that is written as 
\begin{eqnarray}
\hat{H}^{RWA} &=&\hbar \omega _{o}\hat{a}^{\dagger }\hat{a}+\hbar \int
\Omega \hat{b}_{\Omega }^{\dagger }\hat{b}_{\Omega }d\Omega  \label{237} \\
&&\qquad \qquad +\hbar \int \left[ v\left( \Omega \right) \hat{a}^{\dagger }%
\hat{b}_{\Omega }+v^{*}\left( \Omega \right) \hat{a}\hat{b}_{\Omega
}^{\dagger }\right] d\Omega .  \nonumber
\end{eqnarray}
The diagonalization procedure presented in the sequel is basically a review
of the method presented in \cite{Bar1}. Our goal is to find an operator that
satisfies the eigenoperator equation
\begin{equation}
\left[ \hat{A}_{\omega },\hat{H}^{RWA}\right] =\hbar \omega \hat{A}_{\omega
},  \label{241}
\end{equation}
and therefore has its evolution trivially given by $\hat{A}_{\omega }\left(
t\right) =\hat{A}_{\omega }e^{-i\omega t}$.

The new operator $\hat{A}_{\omega }$ can be written in terms of the operator 
$\hat{a}$ of the system and of the operators $\hat{b}_{\Omega }$ of the
reservoir in the form
\begin{equation}
\hat{A}_{\omega }=\alpha _{\omega }\hat{a}+\int d\Omega \beta _{\omega
,\Omega }\hat{b}_{\Omega }.  \label{242}
\end{equation}
Substituting this expression for $\hat{A}_{\omega }$ as well as (\ref{237})
for $\hat{H}^{RWA}$ in (\ref{241}) and calculating the commutators we have
\begin{equation}
\begin{array}{l}
\omega _{o}\alpha _{\omega }\hat{a}+\alpha _{\omega }\int d\Omega v\left(
\Omega \right) \hat{b}_{\Omega }+\int d\Omega \Omega \beta _{\omega ,\Omega }%
\hat{b}_{\Omega } \\ 
\qquad \qquad +\int d\Omega v^{*}\left( \Omega \right) \beta _{\omega
,\Omega }\hat{a}=\omega \left( \alpha _{\omega }\hat{a}+\int d\Omega \beta
_{\omega ,\Omega }\hat{b}_{\Omega }\right) .
\end{array}
\label{243}
\end{equation}
Now, taking the commutator of this expression with $\hat{a}^{\dagger }$ and $%
\hat{b}_{\Omega }^{\dagger }$, we obtain
\begin{eqnarray}
\omega _{o}\alpha _{\omega }+\int d\Omega v^{*}\left( \Omega \right) \beta
_{\omega ,\Omega } &=&\omega \alpha _{\omega },  \label{244} \\
v\left( \Omega \right) \alpha _{\omega }+\Omega \beta _{\omega ,\Omega }
&=&\omega \beta _{\omega ,\Omega },  \label{245}
\end{eqnarray}
respectively. Imposing 
\begin{equation}
\left[ \hat{A}_{\omega },\hat{A}_{\tilde{\omega}}^{\dagger }\right] =\delta
\left( \omega -\tilde{\omega}\right) ,  \label{241b}
\end{equation}
we have 
\begin{equation}
\alpha _{\omega }\alpha _{\tilde{\omega}}^{*}+\int d\Omega \beta _{\omega
,\Omega }\beta _{\tilde{\omega},\Omega }^{*}=\delta \left( \omega -\tilde{%
\omega}\right) .  \label{246}
\end{equation}
The system of equations (\ref{244}), (\ref{245}) and (\ref{246}) is
identical to the one presented in \cite{Fn}. The solution is given by 
\begin{equation}
\left| \alpha _{\omega }\right| ^{2}=\frac{\left| v\left( \omega \right)
\right| ^{2}}{\left[ \omega -\omega _{o}-F\left( \omega \right) \right]
^{2}+\left[ \pi \left| v\left( \omega \right)\right| ^{2}\right] ^{2}},  \label{247}
\end{equation}
with an arbitrary phase of $\alpha _{\omega }$, and 
\begin{equation}
\beta _{\omega ,\Omega }=\left[ {\mathcal P}\frac{1}{\omega -\Omega }+\frac{%
\omega -\omega _{o}-F\left( \omega \right) }{\left| v\left( \omega \right)
\right| ^{2}}\delta \left( \Omega -\omega \right) \right] v\left( \Omega
\right) \alpha _{\omega },  \label{248}
\end{equation}
where 
\begin{equation}
F\left( \omega \right) ={\mathcal P}\int \frac{\left| v\left( \Omega \right)
\right| ^{2}}{\omega -\Omega }d\Omega ,  \label{249}
\end{equation}
and ${\mathcal P}$ denotes the principal part.

We can calculate the evolution of the operator $\hat{a}$ of the system
expressing it as function of the operators $\hat{A}_{\omega }$. We can write 
$\hat{a}$ as function of $\hat{A}_{\omega }$ in the following way 
\begin{equation}
\hat{a}=\int d\omega f_{\omega }\hat{A}_{\omega }.  \label{251}
\end{equation}
Taking the commutator $\left[ \hat{a},\hat{A}_{\omega }^{\dagger }\right] $,
first using (\ref{242}) and then (\ref{251}), we obtain $f_{\omega }=\alpha
_{\omega }^{*}$. Therefore the evolution of the operator $\hat{a}$ is given
by 
\begin{equation}
\hat{a}\left( t\right) =\int d\omega \alpha _{\omega }^{*}\hat{A}_{\omega
}e^{-i\omega t}.  \label{254}
\end{equation}
Substituting the expression for $\hat{A}_{\omega }$ in this equation and
using (\ref{248}) we obtain 
\begin{eqnarray}
\hat{a}\left( t\right)  &=&\int d\omega \left| \alpha _{\omega }\right|
^{2}e^{-i\omega t}\hat{a}  \nonumber \\
&&+\int d\Omega v\left( \Omega \right) \left\{ \int d\omega \left| \alpha
_{\omega }\right| ^{2}{\mathcal P}\frac{1}{\omega -\Omega }e^{-i\omega
t}\right.   \label{255} \\
&&\qquad \left. +\frac{\left| \alpha _{\Omega }\right| ^{2}}{\left| v\left(
\Omega \right) \right| ^{2}}\left[ \Omega -\omega _{o}-F\left( \Omega
\right) \right] e^{-i\Omega t}\right\} \hat{b}_{\Omega }.  \nonumber
\end{eqnarray}

\subsection{The Hamiltonian without the Rotating Wave Approximation}
Now we will present the diagonalization of the Hamiltonian (\ref{235})
without the RWA. The procedure that we will present is similar to the one
adopted in \cite{Bar2}.

Again we want to find an operator $\hat{A}_{\omega }$ that satisfies (\ref
{241}), with $\hat{H}$ in the place of $\hat{H}^{RWA}$, and (\ref{241b}).
Then we write $\hat{A}_{\omega }$ in the form 
\begin{equation}
\hat{A}_{\omega }=\alpha _{\omega }\hat{a}+\int d\Omega \beta _{\omega
,\Omega }\hat{b}_{\Omega }+\chi _{\omega }\hat{a}^{\dagger }+\int d\Omega
\sigma _{\omega ,\Omega }\hat{b}_{\Omega }^{\dagger }.  \label{272}
\end{equation}

Imposing (\ref{241}) and (\ref{241b}) we obtain (see Appendix A) 
\begin{eqnarray}
\left| \alpha _{\omega }\right| ^{2} &=&\left( \frac{\omega +\omega _{o}}{%
2\omega _{o}}\right) ^{2}\frac{1}{\left| v\left( \omega \right) \right|
^{2}\left[ \pi ^{2}+z ^{2}\left( \omega \right) \right] },  \label{273} \\
\beta _{\omega ,\Omega } &=&\left[ {\mathcal P}\frac{1}{\omega -\Omega }%
+z\left( \omega \right) \delta \left( \omega -\Omega \right) \right] \frac{%
2\omega _{o}}{\omega +\omega _{o}}v\left( \Omega \right) \alpha _{\omega },
\label{274} \\
\chi _{\omega } &=&\frac{\omega -\omega _{o}}{\omega +\omega _{o}}\alpha
_{\omega },  \label{275} \\
\sigma _{\omega ,\Omega } &=&\frac{1}{\omega +\Omega }\frac{2\omega _{o}}{%
\omega +\omega _{o}}v^{*}\left( \Omega \right) \alpha _{\omega },
\label{276}
\end{eqnarray}
where 
\begin{equation}
z\left( \omega \right) =\frac{\omega ^{2}-\omega _{o}^{2}-2\omega
_{o}H\left( \omega \right) }{2\omega _{o}\left| v\left( \omega \right)
\right| ^{2}}  \label{277b}
\end{equation}
and 
\begin{equation}
H\left( \omega \right) =F\left( \omega \right) -G\left( \omega \right) =%
{\mathcal P}\int \frac{\left| v\left( \Omega \right) \right| ^{2}}{\omega
-\Omega }d\Omega -\int \frac{\left| v\left( \Omega \right) \right| ^{2}}{%
\omega +\Omega }d\Omega .  \label{278}
\end{equation}

We can express $\hat{a}$ as a function of $\hat{A}_{\omega }$ and $\hat{A}%
_{\omega }^{\dagger }$ in the following way
\begin{equation}
\hat{a}=\int d\omega \phi _{\omega }\hat{A}_{\omega }+\int d\omega \varphi
_{\omega }\hat{A}_{\omega }^{\dagger }.  \label{279}
\end{equation}
Now taking, again, the commutators $\left[ \hat{a},\hat{A}_{\omega
}^{\dagger }\right] $ and $\left[ \hat{a},\hat{A}_{\omega }\right] $ we
obtain $\phi _{\omega }=\alpha _{\omega }^{*}$ and $\varphi _{\omega }=-\chi
_{\omega }$. Substituting the expression (\ref{272}) for $\hat{A}_{\omega }$
in (\ref{279}) the time evolution of the operator $a$ can be easily written
as 
\begin{eqnarray}
\hat{a}\left( t\right) &=&\int \frac{d\omega }{\pi }\left| L\left( \omega
\right) \right| ^{2}\left\{ A\left( \omega \right) \cos \left( \omega
t\right) \hat{a}-i\left[ B\left( \omega \right) \hat{a}+C\left( \omega
\right) \hat{a}^{\dagger }\right] \sin \left( \omega t\right) \right\} \nonumber \\
&&\qquad \qquad \qquad \qquad+\int 
\frac{d\Omega }{\pi }B_{1}\left( \Omega ;t\right) \hat{b}_{\Omega }+\int 
\frac{d\Omega }{\pi }B_{2}\left( \Omega ;t\right) \hat{b}_{\Omega }^{\dagger
},  \label{280}
\end{eqnarray}
where 
\begin{eqnarray}
A\left( \omega \right) &=&2\omega ,\quad B\left( \omega \right) =\frac{%
\omega ^{2}+\omega _{o}^{2}}{\omega _{o}},\quad C\left( \omega \right) =%
\frac{\omega ^{2}-\omega _{o}^{2}}{\omega _{o}},  \label{281} \\
B_{1}\left( \Omega ;t\right) &=&v\left( \Omega \right) \left\{ \left( \omega
_{o}+\Omega \right) \left[ X\left( \Omega ;t\right) +Z\left( \Omega \right)
e^{-i\Omega t}\right] -iY_{\left( +\right) }\left( \Omega ;t\right) \right\}
,  \label{281b} \\
B_{2}\left( \Omega ;t\right) &=&v^{*}\left( \Omega \right) \left\{ \left(
\omega _{o}-\Omega \right) \left[ X\left( \Omega ;t\right) +Z\left( \Omega
\right) e^{i\Omega t}\right] -iY_{\left( -\right) }\left( \Omega ;t\right)
\right\} ,  \label{281d}
\end{eqnarray}
with 
\begin{eqnarray}
X\left( \Omega ;t\right) &=&{\mathcal P}\int d\omega \frac{2\left| L\left(
\omega \right) \right| ^{2}}{\omega ^{2}-\Omega ^{2}}\omega \cos \left(
\omega t\right) ,  \label{281e} \\
Y_{\left( \pm \right) }\left( \Omega ;t\right) &=&{\mathcal P}\int d\omega 
\frac{2\left| L\left( \omega \right) \right| ^{2}}{\omega ^{2}-\Omega ^{2}}%
\left( \omega ^{2}\pm \omega _{o}\Omega \right) \sin \left( \omega t\right) ,
\label{281f} \\
Z\left( \Omega \right) &=&\frac{\left| L\left( \Omega \right) \right| ^{2}}{%
\left| v\left( \Omega \right) \right| ^{2}}\left[ \frac{\Omega ^{2}-\omega
_{o}^{2}}{2\omega _{o}}-H\left( \Omega \right) \right] ,  \label{281g}
\end{eqnarray}
\begin{equation}
\left| L\left( \omega \right) \right| ^{2}=\frac{2\pi \omega _{o}\left|
v\left( \omega \right) \right| ^{2}}{\left[ \omega ^{2}-\omega
_{o}^{2}-2\omega _{o}H\left( \omega \right) \right] ^{2}+\left[ 2\pi \omega
_{o}\left| v\left( \omega \right) \right| ^{2}\right] ^{2}}.  \label{281h}
\end{equation}
\section{The Model of Coordinate-Coordinate Coupling}

The expressions obtained for the evolution of the operator $\hat{a}\left(
t\right) $, within or without the RWA, remained written in terms of the
coupling function $\left| v\left( \omega \right) \right| ^{2}$. Therefore,
the choice of the function $\left| v\left( \omega \right) \right| ^{2}$ will
determine the dynamics of the damped oscillator. We will choose the function 
$\left| v\left( \omega \right) \right| ^{2}$ by comparing the dissipation
model corresponding to the Hamiltonian (\ref{235}) to the one presented in 
\cite{AmB} that corresponds to the following Hamiltonian 
\begin{eqnarray}
\hat{H} &=&\frac{\hat{p}^{2}}{2M}+V\left( \hat{q}\right) +\sum_{j}\left( 
\frac{\hat{p}_{j}^{2}}{2m_{j}}+\frac{m_{j}\omega _{j}^{2}}{2}\hat{q}%
_{j}^{2}\right)  \label{301} \\
&&\qquad \qquad \qquad \qquad -\sum_{j}C_{j}\hat{q}_{j}\hat{q}+V_{R}\left( 
\hat{q}\right) ,  \nonumber
\end{eqnarray}
where the counter-term $V_{R}\left( \hat{q}\right) $, which cancels the
additional contribution to $V\left( \hat{q}\right) $ due to the coupling of
the system to the reservoir, is given by
\begin{equation}
V_{R}\left( \hat{q}\right) =\sum_{j}\frac{C_{j}^{2}}{2m_{j}\omega _{j}^{2}}%
\hat{q}^{2}.  \label{302}
\end{equation}
The spectral function $J\left( \omega \right) $ is defined by 
\begin{equation}
J\left( \omega \right) =\frac{\pi }{2}\sum_{j}\frac{C_{j}^{2}}{m_{j}\omega
_{j}}\delta \left( \omega -\omega _{j}\right) =\frac{\pi }{2}\frac{g\left(
\omega \right) C_{\omega }^{2}}{m_{\omega }\omega },  \label{303b}
\end{equation}
where we have taken the limit of a continuous spectrum and used $g\left( \omega
\right) $ from (\ref{230}). For ohmic dissipation 
\begin{equation}
J\left( \omega \right) =\left\{ 
\begin{array}{c}
2M\gamma \omega \quad if\quad \omega <\Omega _{c} \\ 
0\quad if\quad \omega >\Omega _{c},
\end{array}
\right.  \label{304}
\end{equation}
where $\Omega _{c}$ is a cutoff frequency, much larger than the natural
frequencies of the motion of the system of interest. But in our calculations we
will conveniently use the Drude form 
\begin{equation}
J\left( \omega \right) =\frac{2M\gamma \omega }{\left( 1+\omega ^{2}/\Omega
_{c}^{2}\right) }.  \label{305}
\end{equation}

We are treating a damped harmonic oscillator so $V\left( \hat{%
q}\right) =1/2M\omega _{o}^{2}\hat{q}^{2}$. Applying the usual definitions
of the operators $\hat{a}$ and $\hat{b}_{j}$, 
\begin{equation}
\hat{a}=\sqrt{\frac{M\omega _{o}}{2\hbar }}\left( \hat{q}+\frac{i}{M\omega
_{o}}\hat{p}\right) ,\quad \hat{b}_{j}=\sqrt{\frac{m_{j}\omega _{j}}{2\hbar }%
}\left( \hat{q}_{j}+\frac{i}{m_{j}\omega _{j}}\hat{p}_{j}\right) ,
\label{41}
\end{equation}
we can rewrite (\ref{301}), initially without the inclusion of the
counter-term $V_{R}\left( \hat{q}\right) $, as 
\begin{eqnarray}
\hat{H} &=&\hbar \omega _{o}\hat{a}^{\dagger }\hat{a}+\sum_{j}\hbar \omega
_{j}\hat{b}_{j}^{\dagger }\hat{b}_{j}  \label{42} \\
&&\quad -\frac{\hbar }{2}\sqrt{\frac{1}{M\omega _{o}}}\left( \hat{a}+\hat{a}%
^{\dagger }\right) \sum_{j}\frac{C_{j}}{\sqrt{m_{j}\omega _{j}}}\left( \hat{b%
}_{j}+\hat{b}_{j}^{\dagger }\right)  \nonumber
\end{eqnarray}
(measuring the energy of the system from the energy of the vacuum). Now we
can use the transformation (\ref{230}) in order to consider a continuous
spectrum for the excitations of the reservoir. The second term on the RHS of
(\ref{42}) becomes 
\begin{equation}
\hat{H}_{Res}=\hbar \int \omega \hat{b}_{\omega }^{\dagger }\hat{b}_{\omega
}d\omega  \label{422}
\end{equation}
and its last term can be written in the following way: 
\begin{equation}
\hat{H}_{Int}=-\frac{\hbar }{2}\sqrt{\frac{1}{M\omega _{o}}}\left( \hat{a}+%
\hat{a}^{\dagger }\right) \int d\omega \sqrt{\frac{g\left( \omega \right) }{%
m_{\omega }\omega }}C_{\omega }\left( \hat{b}_{\omega }+\hat{b}_{\omega
}^{\dagger }\right) .  \label{43}
\end{equation}
Now comparing (\ref{42}-\ref{43}) with (\ref{235}) we see that both
Hamiltonians will be equivalent if we employ
\begin{equation}
v\left( \omega \right) =-\frac{1}{2}\sqrt{\frac{g\left( \omega \right) }{%
M\omega _{o}m_{\omega }\omega }}C_{\omega }.  \label{44}
\end{equation}
Taking the square of (\ref{44}) and comparing it with (\ref{303b}) we obtain 
\begin{equation}
v ^{2}\left( \omega \right)=\frac{1}{2\pi }\frac{J\left( \omega \right) }{%
M\omega _{o}}.  \label{46}
\end{equation}
Adopting the Drude form (\ref{305}) $\left| v\left( \omega
\right) \right| ^{2}$ is given by 
\begin{equation}
\left| v\left( \omega \right) \right| ^{2}=\frac{\gamma \omega }{\pi \omega
_{o}}\frac{1}{\left( 1+\omega ^{2}/\Omega _{c}^{2}\right) },  \label{47}
\end{equation}
which is defined only for $\omega \geq 0$.

Now that we have established the form of $\left| v\left( \omega \right)
\right| ^{2}$ corresponding to the Caldeira-Leggett model \cite{AmB}, 
we can determine $H\left( \omega \right) $
through (\ref{278}). A simple calculation shows that $H\left( \omega \right) 
$ will be given by 
\begin{equation}
H\left( \omega \right) =-\frac{\gamma \Omega _{c}}{\omega _{o}}\frac{1}{%
\left( 1+\omega ^{2}/\Omega _{c}^{2}\right) }.  \label{49}
\end{equation}

We can also diagonalize the Hamiltonian (\ref{301}) considering the
inclusion of the counter-term $V_{R}\left( \hat{q}\right) $ (see Appendix
A). The result is that all the equations (\ref{272}-\ref{281h}) will remain
valid with the following substitution: whenever the function $H\left( \omega
\right) $ appears it should be replaced by 
\begin{equation}
H_{R}\left( \omega \right) =H\left( \omega \right) +\frac{\Delta \omega ^{2}%
}{2\omega _{o}},  \label{410}
\end{equation}
where the frequency shift $\Delta \omega ^{2}$ is defined as \cite{AmB} 
\begin{equation}
\frac{\Delta \omega ^{2}}{2\omega _{o}}=\frac{1}{2\omega _{o}M}%
\sum_{j=1}^{N}\frac{C_{j}^{2}}{m_{j}\omega _{j}^{2}}=2\int d\omega \frac{%
\left| v\left( \omega \right) \right| ^{2}}{\omega }=\frac{\gamma \Omega
_{c}}{\omega _{o}}.  \label{411}
\end{equation}
Whenever a function appears with the sub-index $_{R}$ it means that we are
considering the introduction of the counter-term.

The spectral function (\ref{305}) is appropriate to the description of the
reservoir since we consider $\Omega _{c}\gg \omega _{o}, \gamma $. 
So, in order to simplify and also obtain the exact function associated
to the ohmic dissipation we will take the limit $\Omega _{c}\rightarrow
\infty $ in the expression for $\left| L\left( \omega \right)
\right| _{R}^{2}$. 
To do so, first we consider the renormalized function $\left| L\left(
\omega \right) \right| _{R}^{2}$ given by 
\begin{equation}
\left| L\left( \omega \right) \right| _{R}^{2}=\frac{2\pi \omega _{o}\left|
v\left( \omega \right) \right| ^{2}}{\left[ \omega ^{2}-\omega
_{o}^{2}-2\omega _{o}H_{R}\left( \omega \right) \right] ^{2}+\left[ 2\pi
\omega _{o}\left| v\left( \omega \right) \right| ^{2}\right] ^{2}}.
\label{421}
\end{equation}
Once 
\begin{equation}
\lim_{\Omega _{c}\rightarrow \infty }H_{R}\left( \omega \right) =0\text{,}%
\quad \text{and}\quad \lim_{\Omega _{c}\rightarrow \infty }\left| v\left(
\omega \right) \right| ^{2}=\frac{\gamma \omega }{\pi \omega _{o}}\text{,}
\end{equation}
we obtain  
\begin{equation}
\lim_{\Omega _{c}\rightarrow \infty }\left| L\left( \omega \right) \right|
_{R}^{2}=\frac{2\gamma \omega }{\left( \omega ^{2}-\omega _{o}^{2}\right)
^{2}+\left( 2\gamma \omega \right) ^{2}}.
\end{equation}
Thus, we see that 
\begin{equation}
\lim_{\Omega _{c}\rightarrow \infty }\left| L\left( \omega \right) \right|
_{R}^{2}=M\chi ^{"}\left( \omega \right) ,
\end{equation}
where $\chi ^{"}\left( \omega \right) $ is the imaginary part of the
response function of a damped harmonic oscillator.

In the limit $\gamma \ll \omega _{o}$ we can write 
\begin{eqnarray}
\left| L\left( \omega \right) \right| _{R}^{2} &\simeq &\frac{2\omega
_{o}\gamma }{\left[ 2\omega _{o}\left( \omega -\omega _{o}\right) \right]
^{2}+\left( 2\omega _{o}\gamma \right) ^{2}}  \nonumber \\
&=&\frac{1}{2\omega _{o}}\frac{\gamma }{\left( \omega -\omega _{o}\right)
^{2}+\gamma ^{2}},
\end{eqnarray}
that corresponds to a Lorentzian distribution of width $\gamma $.

For the function $\left| L\left( \omega \right) \right| ^{2}$, without the
renormalization, we have $H\left( \omega \ll \Omega _{c}\right) \simeq
-\gamma \Omega _{c}/\omega _{o}$ and therefore for $\Omega _{c}\gg \omega _{o}, \gamma $ we obtain 
\begin{equation}
\left| L\left( \omega \right) \right| ^{2}=\frac{2\gamma \omega }{\left(
\omega ^{2}-\omega _{o}^{2}+2\gamma \Omega _{c}\right) ^{2}+\left( 2\gamma
\omega \right) ^{2}}.  \label{430}
\end{equation}
In this case we should have $\omega _{o}^{2}>2\gamma \Omega _{c}$, because,
without the renormalization, we must have \cite{Bar2}
\begin{equation}
\omega _{o}^{2}>\left| \Delta \omega ^{2}\right| 
\end{equation}
for the diagonalization to be consistent.

\section{Analysis of the Evolution of \lowercase{$\hat{a}\left( t\right) $}}
Now we can analyze in detail the time evolution of the operator $\hat{a}$
associated with the system. We will analyze each term of the
expression for $\hat{a}\left( t\right) $ in eq.(\ref{280}). We will be
interested in the relation between the degree of dissipation in
our system and the importance of each one of those terms.

Initially we will analyze the coefficients associated to the operators $\hat{%
a}$ and $\hat{a}^{\dagger }$. The fastest and most efficient way to
understand the behavior of each one of them is through graphs.

The graphs in Fig.1(a) present the behavior of $\left| L\left( \omega \right)
\right| _{R}^{2}$ for three values of $\gamma $: $\gamma _{1}=0.1\omega _{o}$, $%
\gamma _{2}=\omega _{o}$ and $\gamma _{3}=10\omega _{o}$. We see that for $\gamma _{1}
=0.1\omega _{o}$, $\left| L\left( \omega \right) \right| _{R}^{2}$ presents
a narrow peak centered approximately about $\omega _{o}$ (we showed that in
the limit $\gamma \ll \omega _{o}$ the function $\left| L\left( \omega
\right) \right| _{R}^{2}$ tends to a Lorentzian centered at $\omega _{o}$
and with width $\gamma $). As $\gamma $ increases ($\gamma _{2}=\omega _{o}$)
the function $\left| L\left( \omega \right) \right| _{R}^{2}$ broadens 
and becomes centered at progressively lower frequencies. For $\gamma $ still larger ($%
\gamma _{3}=10\omega _{o}$) $\left| L\left( \omega \right) \right| _{R}^{2}$
narrows again, but its peak is about very low frequencies.

The graphs in Fig.1(b) present the behavior of the functions $A\left( \omega
\right) $, $B\left( \omega \right) $ and $C\left( \omega \right) $ that
appear multiplying $\left| L\left( \omega \right) \right| _{R}^{2}$ in the
different terms of the expression for $\hat{a}\left( t\right) $.
Simultaneously observing (1.a) and (1.b) we conclude that when $\gamma \ll
\omega _{o}$ the function $C\left( \omega \right) \left| L\left( \omega
\right) \right| _{R}^{2}$ has a negligible amplitude if compared to
the functions $A\left( \omega \right) \left| L\left( \omega \right) \right|
_{R}^{2}$ and $B\left( \omega \right) \left| L\left( \omega \right) \right|
_{R}^{2}$, because in this case $\left| L\left( \omega \right) \right|
_{R}^{2}$ is very sharp and centered at $\omega _{o}$ whereas 
$C\left( \omega _{o}\right) =0$. As $\gamma /\omega _{o}$ increases, $\left|
L\left( \omega \right) \right| _{R}^{2}$ has its peak broadened and moved
away from $\omega _{o}$. The function $C\left( \omega \right) \left| L\left(
\omega \right) \right| _{R}^{2}$ becomes comparable to the others and in the
limit $\gamma \gg \omega _{o}$, it is of the same order of $B\left( \omega
\right) \left| L\left( \omega \right) \right| _{R}^{2}$ whereas $A\left(
\omega \right) \left| L\left( \omega \right) \right| _{R}^{2}$ becomes very
small.

It remains to analyze the coefficients $B_{1,R}\left( \Omega ;t\right) $ and $%
B_{2,R}\left( \Omega ;t\right) $ of $\hat{b}_{\Omega }$ and $\hat{b}_{\Omega
}^{\dagger }$, respectively, in the expression (\ref{280}) for $\hat{a}%
\left( t\right) $. We know that in the limit $\gamma \ll \omega _{o}$ the
function $\left| L\left( \omega \right) \right| _{R}^{2}$ tends to a
Lorentzian centered at $\omega _{o}$ and with width $\gamma $. Therefore,
the function $\left( \omega _{o}-\Omega \right) Z_{R}\left( \Omega \right) $
that appears in the expression (\ref{281d}) for $B_{2,R}\left( \Omega
;t\right) $ is, in this limit, negligible if compared to the function $%
\left( \omega _{o}+\Omega \right) Z_{R}\left( \Omega \right) $ in the expression
(\ref{281b}) for $B_{1,R}\left( \Omega ;t\right) $. The evaluation of $X_{R}\left(
\Omega ;t\right) $ results in 
\begin{eqnarray}
X_{R}\left( \Omega ;t\right) &=&\frac{-\pi}{\left( \Omega ^{2}-\omega ^{^{\prime
}2}+\gamma ^{2}\right) +\left( 2\gamma \omega ^{\prime }\right) ^{2}}\frac{d%
}{dt}\left\{ \left[ \frac{\Omega ^{2}-\omega ^{^{\prime }2}+\gamma ^{2}}{\omega
^{\prime }}\sin \left( \omega ^{\prime }t\right) +2\gamma \cos \left( \omega
^{\prime }t\right) \right] e^{-\gamma t}\right\} \nonumber \\
&&\qquad \qquad \qquad \qquad \qquad \qquad \qquad-\frac{2\gamma \Omega \sin \left(
\Omega t\right) }{\left( \Omega ^{2}-\omega _{o}^{2}\right) +\left( 2\gamma
\Omega \right) ^{2}},
\end{eqnarray}
for $\gamma <\omega _{o}$, where $\omega ^{\prime }=\sqrt{\omega _{o}^{2}-\gamma ^{2}}$. We see that in
the limit $\gamma \ll \omega _{o}$ the function $X_{R}\left( \Omega ;t\right) $
will also be very sharply peaked around $\omega _{o}$. Therefore, the function $%
\left( \omega _{o}-\Omega \right) X_{R}\left( \Omega ;t\right) $ in the
expression (\ref{281d}) for $B_{2,R}\left( \Omega ;t\right) $ is also
negligible if compared to the function $\left( \omega _{o}+\Omega \right)
X_{R}\left( \Omega ;t\right) $ in the expression (\ref{281b}) for $B_{1,R}\left(
\Omega ;t\right) $. Similarly it can be shown that, in this limit, the function 
$Y_{\left( -\right),R }\left( \Omega ;t\right) $ is negligible in relation to
the function $Y_{\left( +\right),R }\left( \Omega ;t\right) $. We conclude
that in the limit $\gamma \ll \omega _{o}$ the coefficient $B_{2,R}\left(
\Omega ;t\right) $ is negligible in comparison to the coefficient $%
B_{1,R}\left( \Omega ;t\right) $. As the ratio $\gamma /\omega _{o}$ increases
and the function $\left| L\left( \omega \right) \right| _{R}^{2}$ changes
its shape, the coefficient $B_{2,R}\left( \Omega ;t\right) $ becomes
comparable to $B_{1,R}\left( \Omega ;t\right) $.

So far we have analyzed the relevance of the terms associated to $\hat{a}%
^{\dagger }$ and $\hat{b}_{\Omega }^{\dagger }$ in the expression (\ref{280}%
) for $\hat{a}\left( t\right) $ considering the inclusion of the
counter-term $V_{R}\left( \hat{q}\right) $ in our model. We showed that
these terms are negligible in the limit $\gamma \ll \omega _{o}$, but become
important as the dissipation increases and the function $\left| L\left(
\omega \right) \right| _{R}^{2}$ becomes broader and is no more centered at $%
\omega _{o}$. Now if we had not considered the inclusion of the
counter-term in the interaction Hamiltonian, 
we would have $\left| L\left( \omega \right) \right| ^{2}$ given
by (\ref{430}) instead of $\left| L\left( \omega \right) \right| _{R}^{2}$.
In this case, we see that the condition for the function $\left| L\left(
\omega \right) \right| ^{2}$ to be centered very close to $\omega _{o}$ is
that $2\gamma \Omega _{c}\ll \omega _{o}^{2}$ or 
\begin{equation}
\frac{\gamma }{\omega _{o}}\ll \frac{\omega _{o}}{\Omega _{c}}\ \ \left( \ll
1\right) .  \label{431}
\end{equation}
Therefore, the condition $\gamma /\omega _{o}\ll 1$ would not be enough for
us to ignore the terms associated to $\hat{a}^{\dagger }$ and $\hat{b}%
_{\Omega }^{\dagger }$ in the expression for $\hat{a}\left( t\right) $. 
These terms can only be neglected if the condition (\ref{431}), which limits our system to
a much weaker dissipation, is satisfied.

We notice that a system subject to a weak
dissipation ($\gamma \ll \omega _{o}$, in our case) does not guarantee that
its frequency shift ($\Delta \omega ^{2}=2\gamma \Omega _{c}$)
is also small. We will see later, in more detail, that for a system subject
to very weak dissipation the damping coefficient $\gamma $ will be given by $\pi
\left| v\left( \omega _{o}\right) \right| ^{2}$ and the frequency shift by $%
H\left( \omega _{o}\right) $. Observing the expression (\ref{278}) for $%
H\left( \omega \right) $ we clearly see that the relation between these
functions depends on the form adopted for the function $\left| v\left(
\omega \right) \right| ^{2}$. Therefore, $\pi \left| v\left( \omega
_{o}\right) \right| ^{2}\ll \omega _{o}$ does not guarantee that we will
have $H\left( \omega _{o}\right) \ll \omega _{o}$ (as we have seen to be the case
for $\left| v\left( \omega \right) \right| ^{2}$ given by (\ref{47})),
although this can happen for some functions $\left| v\left( \omega \right)
\right| ^{2}$.

\section{Reduction to the Model with the Rotating Wave Approximation}

Now let us consider the situation in which the following conditions are
satisfied 
\begin{eqnarray}
\pi \left| v\left( \omega \right) \right| ^{2} &\ll &\omega _{o}\text{,\quad
for\quad }\omega \sim \omega _{o}\text{,}  \label{436} \\
H\left( \omega \right) &\ll &\omega _{o}\text{,\quad for\quad }\omega \sim
\omega _{o}\text{.}  \label{433}
\end{eqnarray}
Under these conditions the function $\left| L\left( \omega \right) \right|
^{2}$ will be a function well peaked around $\omega _{o}$. Therefore we can
ignore the terms associated with $\hat{a}^{\dagger }$ and $\hat{b}_{\Omega
}^{\dagger }$ in the expression for $\hat{a}\left( t\right) $. Even the
expressions for the coefficients of $\hat{a}$ and $\hat{b}_{\Omega }$ can be
approximated considering that $\left| L\left( \omega \right) \right| ^{2}$
will only be appreciable, in this case, for $\omega \simeq \omega _{o}$. We
can write 
\begin{equation}
A\left( \omega \right) \left| L\left( \omega \right) \right| ^{2}\simeq
B\left( \omega \right) \left| L\left( \omega \right) \right| ^{2}\simeq
2\omega _{o}\left| L\left( \omega \right) \right| ^{2},
\end{equation}
\begin{eqnarray}
B_{\Omega }^{\left( 1\right) } &\simeq &v\left( \Omega \right) \left\{ \int
d\omega 2\left| L\left( \omega \right) \right| ^{2}{\mathcal P}\frac{%
\omega _{o}}{ \omega -\Omega }e^{-i\omega t}\right.   \nonumber
\\
&&\left. +2\omega _{o}\frac{\left| L\left( \Omega \right) \right| ^{2}}{%
\left| v\left( \Omega \right) \right| ^{2}}\left[ \Omega -\omega
_{o} -H\left( \Omega \right) \right] e^{-i\Omega t}\right\} 
\end{eqnarray}
and finally 
\begin{eqnarray}
\hat{a}\left( t\right)  &=&\int d\omega \left| \tilde{\alpha}_{\omega
}\right| ^{2}e^{-i\omega t}\hat{a}  \nonumber \\
&&+\int d\Omega v\left( \Omega \right) \left\{ \int d\omega \left| \tilde{%
\alpha}_{\omega }\right| ^{2}{\mathcal P}\frac{1}{\omega -\Omega }e^{-i\omega
t}\right.   \label{438} \\
&&\quad \left. +\frac{\left| \tilde{\alpha}_{\Omega }\right| ^{2}}{\left|
v\left( \Omega \right) \right| ^{2}}\left[ \Omega -\omega _{o}-H\left(
\Omega \right) \right] e^{-i\Omega t}\right\} \hat{b}_{\Omega },  \nonumber
\end{eqnarray}
where the function $\left| \tilde{\alpha}_{\omega }\right| ^{2}$ comes from
the approximation of $\left| L\left( \omega \right) \right| ^{2}$
considering (\ref{436} - \ref{433}), 
\begin{equation}
\frac{2\omega _{o}}{\pi }\left| L\left( \omega \right) \right| ^{2}\simeq 
\frac{\left| v\left( \omega \right) \right| ^{2}}{\left[ \omega -\omega
_{o}-H\left( \omega \right) \right] ^{2}+\left[ \pi \left| v\left( \omega
\right) \right| ^{2}\right] ^{2}}=\left| \tilde{\alpha}_{\omega }\right|
^{2}.  \label{432}
\end{equation}

Now let us compare (\ref{438}-\ref{432}) with the expressions (\ref{247})
and (\ref{255}), previously obtained in the RWA. The only difference between
these expressions is given by the presence of $H\left( \omega \right) $
instead of $F\left( \omega \right) $. Once $H\left( \omega \right) -F\left(
\omega \right) =-G\left( \omega \right) $ we would have, for $\omega \sim
\omega _{o}$, $H\left( \omega \right) \simeq F\left( \omega \right) $ if $%
G\left( \omega \right) \ll F\left( \omega \right) $. There can be functions $%
\left| v\left( \omega \right) \right| ^{2}$ that satisfy this requirement.
However, most of the physically reasonable functions $\left| v\left(
\omega \right) \right| ^{2}$ do not; for example, if $\left| v\left( \omega \right)
\right| ^{2}$ is given by (\ref{47}) we have $G\left( \omega \right)
/F\left( \omega \right) \simeq -1$ for $\omega \sim \omega _{o}$. 
In this case, $H\left( \omega _{o}\right) \simeq 2F\left( \omega_{o}\right) $ 
yielding twice the frequency shift given by the model
within the RWA \cite{notabenne}. 
The same relation is found whenever $\left| v\left( \omega \right)
\right| ^{2}$ extends to frequencies much larger than $\omega _{o}$ with
nonnegligible values, for then 
\begin{equation}
H\left( \omega _{o}\right) \simeq -2{\mathcal P}\int \frac{\left| v\left(
\Omega \right) \right| ^{2}}{\Omega }d\Omega \simeq 2F\left( \omega
_{o}\right) .
\end{equation}

This larger frequency shift can be easily
understood through a perturbative analysis. Let us consider a system
described by (\ref{0}), (\ref{1}) and having $\hat{H}_{Int}$ within the RWA (%
\ref{3}). It can be shown that, in second order, the perturbed levels of the
oscillator remain equidistant with an apparent frequency $\omega _{o}+\Delta
^{RWA}\omega $, where \cite{Coh} 
\begin{equation}
\Delta ^{RWA}\omega ={\mathcal P}\sum_{j}\frac{\left| k_{j}\right| ^{2}}{%
\omega _{o}-\omega _{j}}\text{.}  \label{d2}
\end{equation}
Taking the continuous limit and using (\ref{234}) we see that this
expression is nothing but $F\left( \omega _{o}\right) $ which really
represents the frequency shift in the weak dissipation limit. Now it is easy
to show that if we consider $\hat{H}_{Int}$ given by (\ref{2}) without the
RWA, we have in second order in the perturbation, 
\begin{equation}
\Delta \omega ={\mathcal P}\sum_{j}\frac{\left| k_{j}\right| ^{2}}{\omega
_{o}-\omega _{j}}-{\mathcal P}\sum_{j}\frac{\left| k_{j}\right| ^{2}}{\omega
_{o}+\omega _{j}}.
\end{equation}
This expression, in the continuum limit, is nothing but $H\left(
\omega _{o}\right) $. Therefore, we see that the substitution of $F\left(
\omega _{o}\right) $ by $H\left( \omega _{o}\right) $ could already be
foreseen by a simple perturbative theory. The same perturbative analysis can
be used to understand why the counter-rotating term is not important
in the calculation of the decay rate of the system in the weak dissipation
limit. In first order, the decay rate of the system is given by the Fermi's
golden rule for which only the terms of $\hat{H}_{Int}$ that directly
conserve energy in the transition are relevant. This is not done by the
counter-rotating terms. In fact, it is only done by the rotating terms that
create or destroy energy quanta such that $\omega _{j}=\omega _{o}$. This is
the reason for the dependence only on $\left| v\left( \omega _{o}\right) \right|
^{2}$ that appears in the very weak dissipation calculations.

In a model that takes the counter-term into account we automatically
have $H_{R}\left( \omega \right) =0$ and the expression (\ref{438}) can be
substituted by
\begin{eqnarray}
\hat{a}\left( t\right)  &=&\int d\omega \left| \alpha _{\omega }\right|
_{R}^{2}e^{-i\omega t}\hat{a}  \nonumber \\
&&+\int d\Omega v\left( \Omega \right) \left[ \int d\omega \left| \alpha
_{\omega }\right| _{R}^{2}{\mathcal P}\frac{1}{\omega -\Omega }e^{-i\omega
t}\right.  \\
&&\qquad \left. +\frac{\left| \alpha _{\Omega }\right| _{R}^{2}}{\left|
v\left( \Omega \right) \right| ^{2}}\left( \Omega -\omega _{o}\right)
e^{-i\Omega t}\right] \hat{b}_{\Omega },  \nonumber
\end{eqnarray}
where 
\begin{equation}
\frac{2\omega _{o}}{\pi }\left| L\left( \omega \right) \right|
_{R}^{2}\simeq \frac{\left| v\left( \omega \right) \right| ^{2}}{\left(
\omega -\omega _{o}\right) ^{2}+\left[ \pi \left| v\left( \omega \right)
\right| ^{2}\right] ^{2}}=\left| \alpha _{\omega }\right| _{R}^{2}.
\end{equation}

Therefore, the RWA leads us to the correct results, with regard to the decay
rate of the system $\left( \text{related to }\left| \alpha _{\omega }\right|
_{R}^{2}\right) $, {\em if and only if the condition of weak dissipation (\ref
{436}) is satisfied}. Regarding the frequency shift $\left( \text{associated
to }F\left( \omega _{o}\right) \right) $, we see that its agreement with
that given in the limit of weak dissipation, in a model without the
counter-term, strongly depends on the function $\left| v\left( \omega
\right) \right| ^{2}$ adopted. For functions $\left| v\left( \omega \right)
\right| ^{2}$ that extend to frequencies much larger than $\omega _{o}$ we
have twice the shift foreseen in the RWA. Besides, it is also necessary
that the condition (\ref{433}) be satisfied in order to guarantee that this
shift is much smaller than $\omega _{o}$ (and we can neglect the terms in $%
\hat{a}^{\dagger }$ and $\hat{b}_{\Omega }^{\dagger }$ in the expression for 
$\hat{a}\left( t\right) $).

In the case of ohmic dissipation the conditions (\ref{436}-\ref{433}) are
reduced to 
\begin{equation}
\gamma \ll \omega _{o},  \label{433b}
\end{equation}
once in this case 
\[
\pi \left| v\left( \omega _{o}\right) \right| ^{2}=\gamma \text{\quad
and\quad }H_{R}\left( \omega \right) =0\text{,} 
\]
in the limit $\Omega _{c}\rightarrow \infty $.

\section{Evolution of a Coherent State}

We showed that if our system satisfies the conditions of weak dissipation (%
\ref{436}) and small frequency shift (\ref{433}) the evolution of the
operator $\hat{a}\left( t\right) $ can be reduced to the expression given by
(\ref{438}). Now we will suppose that initially our system is in a coherent
state $\left| \alpha \right\rangle $ and that the reservoir is in the vacuum
state $\left| 0\right\rangle $ corresponding to a reservoir at zero
temperature. In this case we have 
\begin{equation}
\hat{a}\left( t\right) \left| \alpha ,0\right\rangle =\int d\omega \left| 
\tilde{\alpha}_{\omega }\right| ^{2}e^{-i\omega t}\alpha \left| \alpha
,0\right\rangle .  \label{257}
\end{equation}
Therefore, in this particular case, a coherent state stays as such
during its evolution with eigenvalue $\alpha \left( t\right) $ given by 
\begin{equation}
\alpha \left( t\right) =\alpha \int d\omega \left| \tilde{\alpha}_{\omega
}\right| ^{2}e^{-i\omega t}.  \label{439b}
\end{equation}
We can also calculate the evolution of the operator $\hat{b}_{\Omega }\left(
t\right) $ of the reservoir. Then in the case of weak dissipation and small
frequency shift we can show that the modes of the reservoir also
evolve from the vacuum state to coherent states with eigenvalues given by 
\begin{eqnarray}
\beta _{\Omega }^{\left( {\mathcal R}\right) }\left( t\right) &=&\alpha \left[ {\mathcal P}
\int d\omega \frac{\left| \tilde{\alpha}_{\omega }\right| ^{2}}{\omega
-\Omega }e^{-i\omega t}\right. \\
&&\qquad \qquad \left. +\frac{\Omega -\omega _{o}-H\left( \Omega \right) }{%
\left| v\left( \Omega \right)\right| ^{2}}\left| \tilde{\alpha}_{\Omega }\right|
^{2}e^{-i\Omega t}\right] v^{*}\left( \Omega \right).  \nonumber
\end{eqnarray}

Still under the conditions (\ref{436}-\ref{433}) we can further 
approximate $\left| \tilde{\alpha}_{\omega }\right|
^{2}$ by 
\begin{equation}
\left| \tilde{\alpha}_{\omega }\right| ^{2}\simeq \frac{\pi \left| v\left(
\omega _{o}\right) \right| ^{2}}{\left[ \omega -\omega _{o}-H\left( \omega
_{o}\right) \right] ^{2}+\left[ \pi \left| v\left( \omega _{o}\right)
\right| ^{2}\right] ^{2}}  \label{435}
\end{equation}
and also extend the lower limit of the frequency integral in (\ref{439b}) to 
$-\infty $ introducing a negligible error. Then we have 
\begin{equation}
\alpha \left( t\right) =\alpha e^{-i\left[ \omega _{o}+\Delta \omega \right]
t}e^{-\pi \left| v\left( \omega _{o}\right) \right| ^{2}t},\ \text{where}\
\Delta \omega =H\left( \omega _{o}\right) \text{.}  \label{441}
\end{equation}
In the case of ohmic dissipation with the inclusion of the counter-term we
have 
\begin{equation}
\alpha \left( t\right) =\alpha e^{-i\omega _{o}t}e^{-\gamma t}.  \label{442}
\end{equation}

Now it is clear that when (\ref{436}-\ref{433}) are not satisfied making 
the terms associated to the operators $\hat{a}^{\dagger }$ and $\hat{%
b}_{\Omega }^{\dagger }$ in the expression (\ref{280}) for $\hat{a}\left(
t\right) $ no longer neglegible, $\left| \alpha ,0\right\rangle $ will not 
be an eigenstate of $%
\hat{a}\left( t\right) $ because $\left| \alpha \right\rangle $ and $\left|
0\right\rangle $ are not eigenstates of $\hat{a}^{\dagger }$ and $\hat{b}%
_{\Omega }^{\dagger }$, respectively. Therefore, we see that an initial
coherent state $\left| \alpha \right\rangle $, interacting with a reservoir
even at temperature $T=0$, will not remain a coherent state during its decay
unless we have a system subject to very weak dissipation.

The previous works that emphasized the existence of dissipative coherent
states \cite{Agar,Wl1,Wl2,Ser}, in models described by the $\hat{H}_{Int}$ (\ref{2}), were based on
master equations obtained through a method that is appropriate only in the limit
of weak dissipation. However, we saw that in this limit the corresponding
model (\ref{2}) is reduced to the RWA model (\ref{3}) that really preserves
the coherent states. We believe that the implicit assumption of weak dissipation is 
the reason why these authors have obtained the dissipative coherent states.
Our result agrees with the one presented in \cite{Zur0} where it was shown
that the model (\ref{301}) presents the coherent states as the initial states of the
system that produce the least amount of entropy as time evolves.

\section{Evolution of the Center of a Wave Packet}

We can also study the evolution of the operator $\hat{q}$ associated to the
position of the particle. Once the operators $\hat{q}$ and $\hat{p}$ are
related to the operator $\hat{a}$ by (\ref{41}), we obtain from (\ref{280})
the following expression for $\hat{q}\left( t\right) $: 
\begin{equation}
\hat{q}\left( t\right) ={\mathcal G}_{\mathcal S}\left( \hat{q},\hat{p};t\right) +%
{\mathcal F}_{\mathcal R}\left( \hat{q}_{\Omega },\hat{p}_{\Omega };t\right) ,
\label{445}
\end{equation}
where 
\begin{eqnarray}
{\mathcal G}_{\mathcal S}\left( \hat{q},\hat{p};t\right) &=&\hat{q}\frac{d}{dt}%
{\mathcal L}\left( t\right) +\frac{\hat{p}}{M}{\mathcal L}\left( t\right) , \\
{\mathcal F}_{\mathcal R}\left( \hat{q}_{\Omega },\hat{p}_{\Omega };t\right)
&=&2\omega _{o}\int \frac{d\Omega }{\pi }v\left( \Omega \right) \sqrt{\frac{%
m_{\Omega }\Omega }{M\omega _{o}}}\left\{ \left[ \frac{d}{dt}W_{R}\left( \Omega
,t\right) +Z_{R}\left( \Omega \right) \cos \left( \Omega t\right) \right] 
\hat{q}_{\Omega }\right.  \nonumber \\
&&\qquad \qquad \qquad \left. +\left[ \Omega W_{R}\left( \Omega
,t\right) +Z_{R}\left( \Omega \right) \sin \left( \Omega t\right) \right]
\right\} \frac{\hat{p}_{\Omega }}{m_{\Omega }\Omega },  \label{531} \\
{\mathcal L}\left( t\right) &=&2\int \frac{d\omega }{\pi }\left| L\left(
\omega \right) \right| _{R}^{2}\sin \left( \omega t\right) , \\
W_{R}\left( \Omega ,t\right) &=&{\mathcal P}\int d\omega \frac{2\left|
L\left( \omega \right) \right| _{R}^{2}}{\omega ^{2}-\Omega ^{2}}\sin \left(
\omega t\right) ,
\end{eqnarray}
with $Z\left( \Omega \right) $ defined in (\ref{281g}).

Now we suppose that the initial density operator of our global system
can be written in the factorizable form 
\begin{equation}
\rho _{T}=\rho _{\mathcal S}\otimes \rho _{\mathcal R},  \label{501}
\end{equation}
where $\rho _{\mathcal S}$ and $\rho _{R}$ are, respectively, the density operators
of the system and reservoir when they are isolated. Then we have 
\begin{eqnarray}
\left\langle \hat{q}\left( t\right) \right\rangle &=&Tr_{\mathcal S}\left[ {\mathcal G}%
_{\mathcal S}\left( \hat{q},\hat{p};t\right) \rho _{\mathcal S}\right] +Tr_{\mathcal R}\left[ {\mathcal F}%
_{\mathcal R}\left( \hat{q}_{\Omega },\hat{p}_{\Omega };t\right) \rho _{\mathcal R}\right] 
\nonumber \\
&=&{\mathcal G}_{\mathcal S}\left( \left\langle \hat{q}\right\rangle _{\mathcal S},\left\langle 
\hat{p}\right\rangle _{\mathcal S};t\right) +{\mathcal F}_{\mathcal R}\left( \left\langle \hat{q%
}_{\Omega }\right\rangle _{\mathcal R},\left\langle \hat{p}_{\Omega }\right\rangle
_{\mathcal R};t\right) .  \label{502}
\end{eqnarray}

Assuming that the initial state of the reservoir is such that 
\begin{equation}
\left\langle \hat{q}_{j}\right\rangle _{\mathcal R}=\left\langle \hat{p}%
_{j}\right\rangle _{\mathcal R}=0,  \label{503}
\end{equation}
which in the continuum limit corresponds to $\left\langle \hat{q}_{\Omega
}\right\rangle _{\mathcal R}=\left\langle \hat{p}_{\Omega }\right\rangle _{\mathcal R}=0$, we 
obtain the following expression for $\left\langle \hat{q}%
\left( t\right) \right\rangle $: 
\begin{equation}
\left\langle \hat{q}\left( t\right) \right\rangle =\left\langle \hat{q}%
\right\rangle _{\mathcal S}\frac{d}{dt}{\mathcal L}\left( t\right) +\frac{\left\langle 
\hat{p}\right\rangle _{\mathcal S}}{M}{\mathcal L}\left( t\right) ,  \label{447}
\end{equation}
where 
\begin{equation}
{\mathcal L}\left( t\right) =\left\{ 
\begin{array}{l}
\frac{1}{\omega ^{\prime }}\sin \left( \omega ^{\prime }t\right) e^{-\gamma
t}, \\ 
t.e^{-\gamma t}, \\ 
\frac{1}{\gamma _{2}-\gamma _{1}}e^{-\gamma _{1}t}+\frac{1}{\gamma
_{1}-\gamma _{2}}e^{-\gamma _{2}t},
\end{array}
\begin{array}{l}
\text{for\qquad }\gamma <\omega _{o}, \\ 
\text{for\qquad }\gamma =\omega _{o}, \\ 
\text{for\qquad }\gamma >\omega _{o},
\end{array}
\right.  \label{506}
\end{equation}
with $\omega ^{\prime }=\sqrt{\omega _{o}^{2}-\gamma ^{2}}$ and $\gamma
_{1,2}=\gamma \pm \sqrt{\gamma ^{2}-\omega _{o}^{2}}$. The expression (\ref
{447}) was also obtained by Grabert and collaborators \cite{Grab}, by the
method of functional integration. They affirmed that it would correspond to
the classical trajectory of a damped harmonic oscillator. However, it is easy
to see that this is not true. If the initial state of the system presents an
initial average momentum $\left\langle \hat{p}\right\rangle _{\mathcal S}=p_{o}$ and
an initial average position $\left\langle \hat{q}\right\rangle _{\mathcal S}=q_{o}$,
then according to (\ref{447}) 
$
\left\langle \hat{q}\left( t\right) \right\rangle  
$ 
would evolve as 
\begin{eqnarray}
\left\langle \hat{q}\left( t\right) \right\rangle  &=&q_{o}\left[ \cos
\left( \omega ^{\prime }t\right) -\frac{\gamma }{\omega ^{\prime }}\sin
\left( \omega ^{\prime }t\right) \right] e^{-\gamma t}  \label{520} \\
&&\qquad \qquad \quad \qquad +\frac{p_{o}}{M\omega ^{\prime }}\sin \left(
\omega ^{\prime }t\right) e^{-\gamma t},  \nonumber
\end{eqnarray}
for $\gamma <\omega _{o}$. However the classical trajectory is known to be 
\begin{eqnarray}
q\left( t\right) _{clas} &=&q_{o}\left[ \cos \left( \omega ^{\prime
}t\right) +\frac{\gamma }{\omega ^{\prime }}\sin \left( \omega ^{\prime
}t\right) \right] e^{-\gamma t}  \label{521} \\
&&\qquad \qquad \quad \qquad +\frac{p_{o}}{M\omega ^{\prime }}\sin \left(
\omega ^{\prime }t\right) e^{-\gamma t}.  \nonumber
\end{eqnarray}
Thus, we see that there is a phase difference between (\ref{520}) and (\ref
{521}) if the oscillator has an initial displacement $q_{o}$.

Let us now suppose that the initial state of the reservoir is such that 
\begin{equation}
\left\langle \hat{q}_{j}\right\rangle _{\mathcal R}=\frac{C_{j}}{m_{j}\omega _{j}^{2}}%
\left\langle \hat{q}\right\rangle _{\mathcal S},\qquad \left\langle \hat{p}%
_{j}\right\rangle _{\mathcal R}=0.  \label{523}
\end{equation}
We can write the expression (\ref{531}) in the discrete limit, replace (\ref
{503}) by (\ref{523}) and return to the continuum limit. Then we obtain (see
Appendix B) the following expression for $\left\langle \hat{q}\left(
t\right) \right\rangle $: 
\begin{equation}
\left\langle \hat{q}\left( t\right) \right\rangle =\left\langle \hat{q}%
\right\rangle _{\mathcal S}\left[ \frac{d}{dt}{\mathcal L}\left( t\right) +2\gamma 
{\mathcal L}\left( t\right) \right] +\frac{\left\langle \hat{p}\right\rangle
_{\mathcal S}}{M}{\mathcal L}\left( t\right) .  \label{524}
\end{equation}
In this case if the initial state of the system presents an initial average
momentum $\left\langle \hat{p}\right\rangle _{\mathcal S}=p_{o}$ and an initial
average position $\left\langle \hat{q}\right\rangle _{\mathcal S}=q_{o}$, (\ref{524})
becomes 
\begin{eqnarray}
\left\langle \hat{q}\left( t\right) \right\rangle  &=&q_{o}\left[ \cos
\left( \omega ^{\prime }t\right) +\frac{\gamma }{\omega ^{\prime }}\sin
\left( \omega ^{\prime }t\right) \right] e^{-\gamma t}  \label{525} \\
&&\quad \qquad \qquad \qquad +\frac{p_{o}}{M\omega ^{\prime }}\sin \left(
\omega ^{\prime }t\right) e^{-\gamma t},  \nonumber
\end{eqnarray}
for $\gamma <\omega _{o}$, that corresponds to the correct classical
trajectory.

Thus, we see that the classical evolution is not obtained with the initial
condition (\ref{503}) but with the initial condition (\ref{523}). We can
understand why this happens through the classical analysis of the model (\ref{305}) 
presented in the next section. 

\section{Classical Analysis and Discussion }

In this section we will accomplish a classical analysis of the model used.
Our objective is to obtain a physical intuition on the effect that causes
the difference between the equations (\ref{520}) and (\ref{525}) and then on the
meaning of the initial condition (\ref{523}). This procedure can be justified by
the equivalence of the classical and quantum dynamics of this model \cite{Hab}%
. 

The Hamiltonian (\ref{301}) can be written as \cite{Hak}: 
\begin{equation}
H=\frac{p^{2}}{2M}+V\left( q\right) +\sum_{j}\left[ \frac{p_{j}^{2}}{2m_{j}}+%
\frac{m_{j}\omega _{j}^{2}}{2}\left( q_{j}-\frac{C_{j}}{m_{j}\omega _{j}^{2}}%
q\right) ^{2}\right] .  \label{306}
\end{equation}
The equations of motion of this system are given by 
\begin{eqnarray}
M\ddot{q}\left( t\right) +V^{\prime }\left( q\right) &=&\sum_{j}C_{j}\left[
q_{j}\left( t\right) -\frac{C_{j}}{m_{j}\omega _{j}^{2}}q\left( t\right)
\right] ,  \label{307} \\
m_{j}\ddot{q}_{j}\left( t\right) +m_{j}\omega _{j}^{2}q_{j}\left( t\right)
&=&C_{j}q\left( t\right) .  \label{308}
\end{eqnarray}
If $q_{j}\left( 0\right) $ and $\dot{q}_{j}\left( 0\right) $ are the initial
conditions the solution of the homogeneous part of (\ref{308}) will be 
\begin{equation}
q_{j}^{H}\left( t\right) =q_{j}\left( 0\right) \cos \left( \omega
_{j}t\right) +\frac{\dot{q}_{j}\left( 0\right) }{\omega _{j}}\sin \left(
\omega _{j}t\right) .  \label{309}
\end{equation}
The particular solution, considering the presence of the force $C_{j}q\left(
t\right) $, can be obtained by taking the Fourier transform of (\ref{308}).
Then we have 
\begin{equation}
\begin{array}{l}
q_{j}^{P}\left( t\right) =\frac{C_{j}}{m_{j}\omega _{j}}\int_{0}^{t}dt^{%
\prime }q\left( t^{\prime }\right) \sin \left[ \omega _{j}\left( t-t^{\prime
}\right) \right]  \\ 
\qquad \,\,=\frac{C_{j}}{m_{j}\omega _{j}^{2}}\left\{ q\left( t\right)
-q\left( 0\right) \cos \left( \omega _{j}t\right) \right.  \\ 
\qquad \qquad \qquad \qquad \left. -\int_{0}^{t}dt^{\prime }\dot{q}\left(
t^{\prime }\right) \cos \left[ \omega _{j}\left( t-t^{\prime }\right)
\right] \right\} 
\end{array}
\label{310}
\end{equation}
Using the definition of the spectral function $J\left( \omega \right) $ (\ref{303b}) 
and (\ref{304}) it can be shown that in the limit $\Omega
_{c}\rightarrow \infty $ we have 
\begin{equation}
\sum_{j}\frac{C_{j}^{2}}{m_{j}\omega _{j}^{2}}\int_{0}^{t}dt^{\prime }\cos
\left[ \omega _{j}\left( t-t^{\prime }\right) \right]\dot{q}\left( t^{\prime }\right)
=2M\gamma \dot{q}\left( t\right) .  \label{312}
\end{equation}
Therefore, the general solution of (\ref{308}), $q_{j}\left( t\right)
=q_{j}^{H}\left( t\right) +q_{j}^{P}\left( t\right) $, when substituted in (%
\ref{307}) results in the following Langevin equation 
\[
M\ddot{q}\left( t\right) +V^{\prime }\left( q\right) +2M\gamma \dot{q}\left(
t\right) =F\left( t\right) , 
\]
where 
\begin{equation}
F\left( t\right) =\sum_{j}C_{j}\tilde{q}_{j}\left( 0\right) \cos \left(
\omega _{j}t\right) +\sum_{j}\frac{C_{j}}{\omega _{j}}\dot{q}_{j}\left(
0\right) \sin \left( \omega _{j}t\right) ,  \label{313}
\end{equation}
is the fluctuating force and we have redefined the position of the
oscillators of the bath \cite{nota}
\begin{equation}
\tilde{q}_{j}\left( 0\right) =q_{j}\left( 0\right) -\frac{C_{j}}{m_{j}\omega
_{j}^{2}}q\left( 0\right) .  \label{314}
\end{equation}
Supposing that the bath is initially in thermodynamic equilibrium in
relation to the coordinates $\tilde{q}_{j}\left( 0\right) $ we have, in the
classical limit, 
\begin{eqnarray}
\left\langle \tilde{q}_{j}\left( 0\right) \right\rangle  &=&\left\langle 
\dot{q}_{j}\left( 0\right) \right\rangle =\left\langle \tilde{q}_{j}\left(
0\right) \dot{q}_{j^{\prime }}\left( 0\right) \right\rangle =0.  \label{332} \\
\left\langle \tilde{q}_{j}\left( 0\right) \tilde{q}_{j^{\prime }}\left(
0\right) \right\rangle  &=&\frac{kT}{m_{j}\omega _{j}^{2}}\delta
_{jj^{\prime }},\ \left\langle \dot{q}_{j}\left( 0\right) \dot{q}_{j^{\prime
}}\left( 0\right) \right\rangle =\frac{kT}{m_{j}}\delta _{jj^{\prime }}.
\label{333}
\end{eqnarray}
The physical meaning of this initial condition written in terms of the 
relative coordinates $\tilde{q}_{j} $ has alread been analized by
Zwanzig \cite{Zwan} some time ago. Using (\ref{332}-\ref{333}) and after some algebraic manipulations it is
shown that $\left\langle F\left( t\right) \right\rangle =0$ and $%
\left\langle F\left( t\right) F\left( t^{\prime }\right) \right\rangle
\simeq 4M\gamma kT\delta \left( t-t^{\prime }\right) $ which correspond to
the expressions that characterize the Brownian motion.

On the other hand if we had adopted the initial condition 
\begin{equation}
\left\langle q_{j}\left( 0\right) \right\rangle =\left\langle \dot{q}%
_{j}\left( 0\right) \right\rangle =\frac{1}{m_{j}}\left\langle p_{j}\left(
0\right) \right\rangle =0,  \label{511}
\end{equation}
we would have 
\begin{eqnarray}
\left\langle F\left( t\right) \right\rangle &=&-q\left( 0\right)\sum_{j}\frac{C_{j}^{2}%
}{m_{j}\omega _{j}^{2}}\cos \left( \omega _{j}t\right)  \nonumber \\
&=&-4M\gamma q\left( 0\right)\frac{1}{\pi }\int_{0}^{\Omega _{c}}d\omega \cos \omega
t=-4M\gamma q\left( 0\right)\delta \left( t\right) ,  \label{512}
\end{eqnarray}
where we have used (\ref{303b}), (\ref{304}) and taken the limit $\Omega
_{c}\rightarrow \infty $. Therefore, we would not have $\left\langle F\left(
t\right) \right\rangle =0$, but the presence of a delta force at $t=0$.
Physically what happens is that if the oscillators of the bath are not
``appropriately'' distributed around the particle (as in the initial
condition (\ref{511})), when it is inserted in the bath, these oscillators
will ``pull'' the particle until they reach this ``appropriate''
distribution. This force will act on the particle during a time interval of
the order $1/\Omega _{c}$. Therefore, in the limit $\Omega _{c}\rightarrow
\infty $ we will have a delta force that will cause a phase difference in
the evolution of the system. This phase difference is the difference between
(\ref{520}) and (\ref{521}) which is corrected in (\ref{525}) by the adoption of the
initial condition (\ref{523}) (quantum analogue of (\ref{332})) instead of (\ref{503})
(quantum analogue of (\ref{511})). As far as we know, the need to use the initial
condition (\ref{523}) in place of (\ref{503}) in the quantum treatment of this
model has not been noticed in previous works.
In Ref.~\cite{AmB} the authors make some approximations which are
equivalent to regarding the initial time as $t=0^{+}$ $\left( t\sim 1/\Omega
_{c}\right) $. So the initial conditions are established at
this instant although the coupling between particle and bath is switched on
at $t=0$ and gives rise to a delta type force at this instant. The inclusion
of $t=0$ in propagator methods must be accompanied by the above-mentioned
modification of the factorizable initial condition. However, it must be
emphasized that we are not addressing here the question of the generalized
initial condition \cite{Grab,Cris}. Actually the point we have
raised is clearly responsible for the disagreement between $\left\langle 
\hat{q}\left( t\right) \right\rangle $ found in these references. In 
Ref.~\cite{Grab} the authors reproduced the dephased $\left\langle \hat{q}\left(
t\right) \right\rangle $ (c.f.\ eq.~(\ref{447}) above) whereas in 
Ref.~\cite{Cris} this time evolution is the correct one as in (\ref{525}). The origin
of the discrepancy is the use of $t=0$ or $t=0^{+}$ as the initial instant
together with the factorizable initial condition.

We would like to take advantage of this opportunity to correct a mistake
that was made in Ref.~\cite{Cris} of which one of us is co-author. The
referred article considers an initial condition of the system when the bath
of oscillators meets thermodynamical equilibrium with the particle at the
position it is placed in the bath. In this case one obtains mean
values of the position $\left\langle \hat{q}\left( t\right) \right\rangle $ and
momentum $\left\langle \hat{p}\left( t\right) \right\rangle $ which depend on the
temperature of the reservoir and that do not exactly coincide with their
classical counterparts. This disagreement was justified within a classical
analysis of the model. In this analysis it was affirmed that the classical
initial condition equivalent to the proposed quantum initial state, that
corresponds exactly to $\left\langle \tilde{q}_{j}\left( 0\right)
\right\rangle =0$, would imply in a classical solution of the model
different from the trajectories of a damped harmonic oscillator. We saw in
the present work that this is not true and therefore this argument can not
be used. We believe that the origin of the disagreement when
adopting a non-factorizable initial condition is the impossibility to
describe the evolution of the system through an independent sum of functions of the
system and reservoir variables as in (\ref{502}). The quantum effects of the
correlation between the variables of the system and reservoir prevent a
direct comparison of the quantum mean values with the values obtained
through the classical analysis of the model. Accordingly, it can be shown that
the discrepancy vanishes in the classical limit $\left( kT\gg \hbar \omega
_{o}\right) $.

After we had made the above analysis we became aware that in previous
works \cite{Pab,Zur} the authors had also noticed the existence of initials kicks and
jolts in this system when the initial condition (\ref{501},\ref{503}) is used. In both
works
the existence of an initial kick, given by (\ref{512}) in the limit limite $%
\Omega _{c}\rightarrow \infty $, is noticed (eqs. (3.2) and (45),
respectively). However, the existence of this initial transient is
considered as a characteristic of the model to be taken into account. In our
analysis we see that although the existence of the kick given by (\ref{512}) is a
real characteristic of the model when it is subject to the initial condition
(\ref{501},\ref{503}) it is an undesirable feature that should be corrected.
Fortunately, this correction can be made even with an improved
factorizable initial
condition, that is, (\ref{501},\ref{523}).

The authors of \cite{Pab} and \cite{Zur} have recognized that the presence of
initial jolts, in their master equation coefficients, generates 
certain non-physical effects and so they suppose
that they are due the adoption of a factorizable initial condition. In \cite
{Pab2} the evolution of the system is analyzed for a non-factorizable
initial condition, similar to the one used in \cite{Grab} and \cite{Cris}, in
which the initial position of the particle is defined by a measurement
process in a state in thermodynamic equilibrium with the
bath. However, the initial jolts in the time scale $1/\Omega _{c}$ still persists. We believe that this
happens because this initial condition does not satisfy (\ref{523}) when the
initial mean values of the position of the particle and the oscillators
in the bath are calculated. Thus, in this aspect, it is less general than
the improved factorizable initial condition that we considered.

Actually the initial jolt, at least the one they attribute to the
decoherence process (however see discussion below), does not appear in the
more general initial condition later adopted in \cite{Zur2}. It is an initial
condition prepared by a dynamic process in a finite time $t_{p}$. In this
case we can consider that the condition (\ref{523}) will be satisfied since $%
t_{p}\gg 1/\Omega _{c}$. Indeed in this situation $\left( t_{p}\gg 1/\Omega
_{c}\right) $, it was shown that the initial jolt does not appear.

Thus, we believe that the initial condition (\ref{523}) is enough to eliminate
most of or maybe all the initial transients that would appear in this system
in the characteristic time scale $1/\Omega _{c}$. However, another initial
transient in this system is also known. In \cite{Amb} it was shown that for a
factorizable initial state in the high temperature limit $\left( kT\gg \hbar
\Omega _{c}\right) $ of the master equation it presents an initial transient
within the time scale of the internal decoherence of the initial wave packet. If
applied to times shorter than this it can lead to nonsensical results. We
believe that this pathology can only be really corrected with the adoption
of non-factorizable initial conditions.

\section{Conclusion}

In this paper we have applied the Fano diagonalization procedure to two
Hamiltonians commonly used as models for dissipative systems in
quantum optics and in condensed matter systems; the rotating wave and the
coordinate-coordinate coupling models, respectively.

By exactly diagonalizing these two models we have succeeded in showing how
the RWA turns out to be the extremely underdamped limit of the more general
coordinate-coordinate coupling model. We have also been able to analyze the
role played by the counter-term in this limiting procedure from the latter
to the RWA. We have shown through the evaluation of the destruction operator 
$\hat{a}\left( t\right) $ of the system that the RWA is a good approximation for (%
\ref{0}) if and only if the conditions (\ref{436}) and (\ref{433}) are
satisfied. For certain choices of $\left| v\left( \omega \right) \right|
^{2}$, we have $H\left( \omega _{o}\right) \approx \pi \left| v\left(
\omega _{o}\right) \right| ^{2}$ and the fulfillment of (\ref{436})
automatically implies (\ref{433}). However, for other choices, we can have 
$H\left( \omega _{o}\right) \gg \pi \left| v\left( \omega _{o}\right)
\right| ^{2}$ and (\ref{433}) limits the validity of the approximation. Once
these conditions are satisfied, we have shown that the time evolution of the 
system is identical to that determined within the RWA, with the exception of
the frequency shift. We have found that this shift will be given by $H\left(
\omega _{o}\right) $ instead of $F\left( \omega _{o}\right) $. As we have shown 
these functions usually have the same order of magnitude, but they are not
identical. For functions $\left| v\left( \omega \right) \right| ^{2}$ that
extend to frequencies much larger than $\omega _{o}$ we have $H\left(
\omega _{o}\right) \simeq 2F\left( \omega _{o}\right) $.

The comparison of the Hamiltonian (\ref{235}) with the Hamiltonian of the
coordinate-coordinate coupling model established the relation (\ref{46})
between the spectral function $J\left( \omega \right) $ of this model and
the coupling function $\left| v\left( \omega \right) \right| ^{2}$. In the
case of ohmic dissipation and considering the inclusion of the counter-term,
we find that $H_{R}\left(\omega \right) =0$  
in the limit $\Omega _{c}\rightarrow \infty $. 
Then the only condition required for the RWA to be valid is 
\begin{equation}
\gamma \ll \omega _{o}.  \label{53}
\end{equation}

As an application of this method, we have studied the existence
of dissipative coherent states and concluded that it can only exist within
the RWA and when thermal fluctuations are neglegible. When these conditions are 
not met, the initial state will in the long run become a statistical mixture.

Finally, we have also addressed the question of the discrepancies
in the time evolution of the observables of the
system that arise when the factorizable initial 
conditions are not properly accounted for. We have shown how to deal with 
this problem by using the appropriate improved factorizable initial condition 
(\ref{523}) rather than (\ref{503}).

\acknowledgments
M. R. da C. acknowledges full support from CAPES. A. O. C. is grateful for
partial support from the Conselho Nacional de Desenvolvimento Cient\'{i}fico
e Tecnol\'{o}gico (CNPq) and H. W., Jr. kindly acknowledges full support from 
Funda\c{c}\~{a}o de Amparo \`{a} Pesquisa do Estado de S\~{a}o Paulo (FAPESP). 
The research of S.\ M.\ Dutra at UNICAMP has also been made possible by a 
postdoctoral grant from FAPESP.

\appendix
\section{Diagonalization without the RWA}

Here, the procedure used in the diagonalization of the Hamiltonian (\ref{235}%
) will be presented. We want to find the operator $\hat{A}_{\omega }$ that
allows us to write (\ref{235}) in the diagonal form. We write $\hat{A}%
_{\omega }$ in its general form (\ref{272}) and then we impose the
commutation relation (\ref{241}) 
\begin{equation}
\left[ \hat{A}_{\omega },\hat{H}\right] =\hbar \omega \hat{A}_{\omega }.
\label{a3}
\end{equation}

Replacing (\ref{235}) and (\ref{272}) in (\ref{a3}) and taking the
commutators of the expression obtained with $\hat{a}^{\dagger }$, $\hat{a}$, 
$\hat{b}_{\Omega }$ and $\hat{b}_{\Omega }^{\dagger }$, we have,
respectively, 
\begin{eqnarray}
\omega \alpha _{\omega } &=&\omega _{o}\alpha _{\omega }+\int \left[ \beta
_{\omega ,\Omega }v^{*}\left( \Omega \right) -\sigma _{\omega ,\Omega
}v\left( \Omega \right) \right] d\Omega ,  \label{a5} \\
\omega \chi _{\omega } &=&-\omega _{o}\chi _{\omega }+\int \left[ \beta
_{\omega ,\Omega }v^{*}\left( \Omega \right) -\sigma _{\omega ,\Omega
}v\left( \Omega \right) \right] d\Omega ,  \label{a6} \\
\omega \beta _{\omega ,\Omega } &=&\left( \alpha _{\omega }-\chi _{\omega
}\right) v\left( \Omega \right) +\Omega \beta _{\omega ,\Omega },  \label{a7}
\\
\omega \sigma _{\omega ,\Omega } &=&\left( \alpha _{\omega }-\chi _{\omega
}\right) v^{*}\left( \Omega \right) -\Omega \sigma _{\omega ,\Omega }.
\label{a8}
\end{eqnarray}
Subtracting (\ref{a5}) from (\ref{a6}), we have 
\begin{equation}
\chi _{\omega }=\frac{\omega -\omega _{o}}{\omega +\omega _{o}}\alpha
_{\omega }.  \label{a9}
\end{equation}
Replacing (\ref{a9}) in (\ref{a7}) we obtain
\begin{equation}
\beta _{\omega ,\Omega }=\left[ {\mathcal P}\frac{1}{\omega -\Omega }+z\left(
\omega \right) \delta \left( \omega -\Omega \right) \right] \frac{2\omega
_{o}}{\omega +\omega _{o}}v\left( \Omega \right) \alpha _{\omega },
\label{a10}
\end{equation}
where $z\left( \omega \right) $ is a function to be determined. Similarly,
substituting (\ref{a9}) in (\ref{a8}), we have 
\begin{equation}
\sigma _{\omega ,\Omega }=\frac{1}{\omega +\Omega }\frac{2\omega _{o}}{%
\omega +\omega _{o}}v^{*}\left( \Omega \right) \alpha _{\omega }.
\label{a11}
\end{equation}
Now, substituting (\ref{a10}-\ref{a11}) in (\ref{a5}) we obtain $%
z\left( \omega \right) $ given by (\ref{277b}).

It remains to determine $\alpha _{\omega }$. For this we impose the
condition (\ref{241b}) which results in 
\begin{equation}
\begin{array}{l}
\alpha _{\omega }\alpha _{\tilde{\omega}}^{*}+\int d\Omega \beta _{\omega
,\Omega }\beta _{\tilde{\omega},\Omega }^{*}-\chi _{\omega }\chi _{\tilde{%
\omega}}^{*} \\ 
\qquad \qquad \qquad \qquad -\int d\Omega \sigma _{\omega ,\Omega }\sigma _{%
\tilde{\omega},\Omega }^{*}=\delta \left( \omega -\tilde{\omega}\right) .
\end{array}
\label{a15}
\end{equation}
Using (\ref{a9}) and (\ref{a11}) we obtain, respectively, 
\begin{equation}
\alpha _{\omega }\alpha _{\tilde{\omega}}^{*}-\chi _{\omega }\chi _{\tilde{%
\omega}}^{*}=\frac{2\omega _{o}\left( \omega +\tilde{\omega}\right) }{\left(
\omega +\omega _{o}\right) \left( \tilde{\omega}+\omega _{o}\right) }\alpha
_{\omega }\alpha _{\tilde{\omega}}^{*}.  \label{a16}
\end{equation}
and 
\begin{equation}
\int d\Omega \sigma _{\omega ,\Omega }\sigma _{\tilde{\omega},\Omega }^{*}=%
\frac{\left( 2\omega _{o}\right) ^{2}}{\left( \omega +\omega _{o}\right)
\left( \tilde{\omega}+\omega _{o}\right) }\frac{G\left( \tilde{\omega}%
\right) -G\left( \omega \right) }{\omega -\tilde{\omega}}\alpha _{\omega
}\alpha _{\tilde{\omega}}^{*},  \label{a17}
\end{equation}
were $G\left( \omega \right) $ is given by (\ref{278}). Now, using (\ref{a10}%
), as well as the property 
\begin{equation}
\frac{{\mathcal P}}{\omega -\omega ^{\prime }}\frac{{\mathcal P}}{\tilde{\omega%
}-\omega ^{\prime }}=\frac{1}{\omega -\tilde{\omega}}\left( \frac{{\mathcal P}%
}{\tilde{\omega}-\omega ^{\prime }}-\frac{{\mathcal P}}{\omega -\omega
^{\prime }}\right) +\pi ^{2}\delta \left( \omega -\tilde{\omega}\right)
\delta \left[ \omega ^{\prime }-\frac{1}{2}\left( \omega +\tilde{\omega}%
\right) \right] ,
\end{equation}
we obtain 
\begin{eqnarray}
\int d\Omega \beta _{\omega ,\Omega }\beta _{\tilde{\omega},\Omega }^{*}&=&%
\frac{\left( 2\omega _{o}\right) ^{2}}{\left( \omega +\omega _{o}\right)
\left( \tilde{\omega}+\omega _{o}\right) }\left\{ \frac{1}{\omega -\tilde{%
\omega}}\left[ \frac{\tilde{\omega}^{2}-\omega ^{2}}{2\omega _{o}}+G\left( 
\tilde{\omega}\right) -G\left( \omega \right) \right] \right. \nonumber \\
&&\qquad \qquad \qquad \qquad \qquad \left.+\left[ \pi
^{2}+z^{2}\left( \omega \right) \right] \left| v\left( \omega \right)
\right| ^{2}\delta \left( \omega -\tilde{\omega}\right) \right\} .
\label{a18}
\end{eqnarray}
Then substituting (\ref{a16}-\ref{a17}) and (\ref{a18}) in (\ref{a15}), we have 
\begin{equation}
\alpha _{\omega }\alpha _{\tilde{\omega}}^{*}\frac{\left( 2\omega
_{o}\right) ^{2}\left| v\left( \omega \right) \right| ^{2}}{\left( \omega
+\omega _{o}\right) \left( \tilde{\omega}+\omega _{o}\right) }\left[ \pi
^{2}+z^{2}\left( \omega \right) \right] \delta \left( \omega -\tilde{\omega}%
\right) =\delta \left( \omega -\tilde{\omega}\right)  \label{a19}
\end{equation}
and, therefore, we should have $\left| \alpha _{\omega }\right| ^{2}$ given
by (\ref{273}).

In the calculations presented above we supposed that $\left| v\left( \omega
\right) \right| $ is a continuous function and such that $\left| v\left(
0\right) \right| =0$. In this way we guarantee that $\int_{0}^{\infty
}d\Omega f\left( \Omega \right) \left| v\left( \Omega \right) \right|
^{2}\delta \left( \Omega -\omega \right) =f\left( \omega \right) \left|
v\left( \omega \right) \right| ^{2}$ for any nonsingular function $f\left(
\omega \right) $ within the whole interval $\left( 0,\infty \right) $.

We can also diagonalize the Hamiltonian (\ref{301}) considering the
introduction of the counter-term $V_{R}\left( \hat{q}\right) $. Rewriting it
in terms of the operators $\hat{a}$ and $\hat{b}_{j}$, defined in (\ref{41}%
), we have 
\begin{equation}
\hat{H}=\hbar \omega _{o}\hat{a}^{\dagger }\hat{a}+\hbar \frac{\Delta \omega
^{2}}{4\omega _{o}}\left( \hat{a}+\hat{a}^{\dagger }\right)
^{2}+\sum_{j}\hbar \omega _{j}\hat{b}_{j}^{\dagger }\hat{b}_{j}-\frac{\hbar }{2}%
\sqrt{\frac{1}{M\omega _{o}}}\left( \hat{a}+\hat{a}^{\dagger }\right)
\sum_{j}\frac{C_{j}}{\sqrt{m_{j}\omega _{j}}}\left( \hat{b}_{j}+\hat{b}_{j}%
^{\dagger }\right) .  \label{c2}
\end{equation}
Writing (\ref{c2}) in the continuum limit and following the same procedure
as adopted above, we will see that the equations (\ref{a5}) and (\ref{a6})
will be substituted now by the equations 
\begin{eqnarray}
\omega \alpha _{\omega } &=&\left( \omega _{o}+\frac{\Delta \omega ^{2}}{%
2\omega _{o}}\right) \alpha _{\omega }-\frac{\Delta \omega ^{2}}{2\omega _{o}%
}\chi _{\omega }+\int \left[ \beta _{\omega ,\Omega }v^{*}\left( \Omega
\right) -\sigma _{\omega ,\Omega }v\left( \Omega \right) \right] d\Omega ,
\label{c3} \\
\omega \chi _{\omega } &=&-\left( \omega _{o}+\frac{\Delta \omega ^{2}}{%
2\omega _{o}}\right) \chi _{\omega }+\frac{\Delta \omega ^{2}}{2\omega _{o}}%
\alpha _{\omega }+\int \left[ \beta _{\omega ,\Omega }v^{*}\left( \Omega
\right) -\sigma _{\omega ,\Omega }v\left( \Omega \right) \right] d\Omega ,
\label{c4}
\end{eqnarray}
respectively. The equations (\ref{a7}) and (\ref{a8}) will stay the same.
Thus, it can be easily shown that all the other previous equations will not
change with the only difference that the function $H\left( \omega \right) $
should be substituted by $H_{R}\left( \omega \right) $ given in (\ref{410}).

\section{Calculation of \lowercase{$\uppercase{{\mathcal F}}_{\uppercase{\mathcal R}}\left( \left\langle \hat{q}_{j}\right\rangle
_{\uppercase{\mathcal R}},\left\langle \hat{p}_{j}\right\rangle _{\uppercase{\mathcal R}};t\right) $}}
The expression (\ref{531}) for ${\mathcal F}_{\mathcal R}\left( \hat{q}_{\Omega },\hat{%
p}_{\Omega };t\right) $ can be written as 
\begin{equation}
{\mathcal F}_{\mathcal R}\left( \hat{q}_{\Omega },\hat{p}_{\Omega };t\right) =2\omega
_{o}\int \frac{d\Omega }{\pi }\sqrt{\frac{m_{\Omega }\Omega }{M\omega _{o}}}%
v\left( \Omega \right) \left[ {\mathcal J}\left( \Omega ;t\right) \hat{q}%
_{\Omega }+{\mathcal K}\left( \Omega ;t\right) \frac{\hat{p}_{\Omega }}{%
m_{\Omega }\Omega }\right] ,  \label{81}
\end{equation}
where the expressions for ${\mathcal J}\left( \Omega ;t\right) $ and ${\mathcal K}%
\left( \Omega ;t\right) $ are obtained by direct comparison between (\ref{81}%
) and (\ref{531}). Now we can substitute the expression (\ref{44}) for $%
v\left( \Omega \right) $ in (\ref{81}) and write the expression obtained in
the discrete limit 
\begin{eqnarray}
{\mathcal F}_{\mathcal R}\left( \hat{q}_{\Omega },\hat{p}_{\Omega };t\right) &=&-\frac{1%
}{M}\sum_{j}\frac{C_{\Omega _{j}}}{\pi }\left[ {\mathcal J}\left( \Omega
_{j};t\right) \sqrt{g\left( \Omega _{j}\right) }\int_{1/g\left( \Omega
_{j}\right) }d\Omega \hat{q}_{\Omega } \right. \nonumber \\
&&\qquad \qquad \qquad \qquad \left. +\frac{{\mathcal K}\left( \Omega
_{j};t\right) }{m_{\Omega _{j}}\Omega _{j}}\sqrt{g\left( \Omega _{j}\right) }%
\int_{1/g\left( \Omega _{j}\right) }d\Omega \hat{p}_{\Omega }\right] .  \label{83}
\end{eqnarray}
Recalling the relation (\ref{230}) between the discrete and continuous
operators, we obtain 
\begin{equation}
{\mathcal F}_{\mathcal R}\left( \hat{q}_{j},\hat{p}_{j};t\right) =-\frac{1}{M}\sum_{j}%
\frac{C_{_{j}}}{\pi }\left[ {\mathcal J}\left( \Omega _{j};t\right) \hat{q}_{j}+%
{\mathcal K}\left( \Omega _{j};t\right) \frac{\hat{p}_{j}}{m_{\Omega _{j}}\Omega
_{j}}\right] .  \label{84}
\end{equation}
Employing the initial condition (\ref{523}), we have 
\begin{eqnarray}
{\mathcal F}_{\mathcal R}\left( \left\langle \hat{q}_{j}\right\rangle
_{R},\left\langle \hat{p}_{j}\right\rangle _{\mathcal R};t\right) &=&-\frac{1}{M\pi }%
\sum_{j}\frac{C_{j}^{2}}{m_{j}\Omega _{j}^{2}}\ {\mathcal J}\left( \Omega
_{j};t\right) \left\langle \hat{q}\right\rangle _{\mathcal S}  \nonumber \\
&=&{\mathcal H}\left( t\right) \left\langle \hat{q}\right\rangle _{\mathcal S},
\end{eqnarray}
with 
\begin{equation}
{\mathcal H}\left( t\right) =-4\omega _{o}\int \frac{d\Omega }{\pi }\frac{%
\left| v\left( \Omega \right) \right| ^{2}}{\Omega }{\mathcal J}\left( \Omega
;t\right) ,
\end{equation}
where we used again the relation (\ref{44}). Writing ${\mathcal H}\left(
t\right) $ as 
\begin{equation}
{\mathcal H}\left( t\right) =I_{1}\left( t\right) +I_{2}\left( t\right) ,
\label{8102}
\end{equation}
we have 
\begin{eqnarray}
I_{1}\left( t\right) &=&-4\omega _{o}\int \frac{d\Omega }{\pi }\frac{\left|
v\left( \Omega \right) \right| ^{2}}{\Omega }\frac{d}{dt}W_{R}\left( \Omega ,t\right) ,
\label{811} \\
I_{2}\left( t\right) &=&-4\omega _{o}\int \frac{d\Omega }{\pi }\frac{\left|
v\left( \Omega \right) \right| ^{2}}{\Omega }Z_{R}\left( \Omega \right) \cos
\left( \Omega t\right) .
\end{eqnarray}
The calculation of $I_{1}\left( t\right) $ is a somewhat lengthy but
straighforward calculation and results in $I_{1}\left( t\right) =0$. So all
that is left is 
\begin{equation}
{\mathcal H}\left( t\right) =I_{2}\left( t\right) =4\gamma \int \frac{d\Omega 
}{\pi }\frac{\omega _{o}^{2}-\Omega ^{2}}{\left( \Omega ^{2}-\omega
_{o}^{2}\right) ^{2}+\left( 2\gamma \Omega \right) ^{2}}\cos \left( \Omega
t\right) .  \label{813}
\end{equation}
The evaluation of this last integral can also be accomplished by the method
of residues and yields 
\begin{equation}
{\mathcal H}\left( t\right) =2\gamma {\mathcal L}\left( t\right) ,  \label{814}
\end{equation}
for $t>0$.

\bibliographystyle{plain}

\begin{thebibliography}{99}
\bibitem{notabenne} However, this frequency shift is still much smaller than $\omega _{o}$.
\bibitem{nota} The distribution $\tilde{q}_{j} =0$ is the one which minimizes the potential energy of the
global system (system plus reservoir) for a fixed value of $q$, as we can see directly from (\ref{306}).
\end{thebibliography}

\begin{figure}
\vbox to 5.7cm {\vss\hbox to 8.2cm
 {\hss\
   {\includegraphics{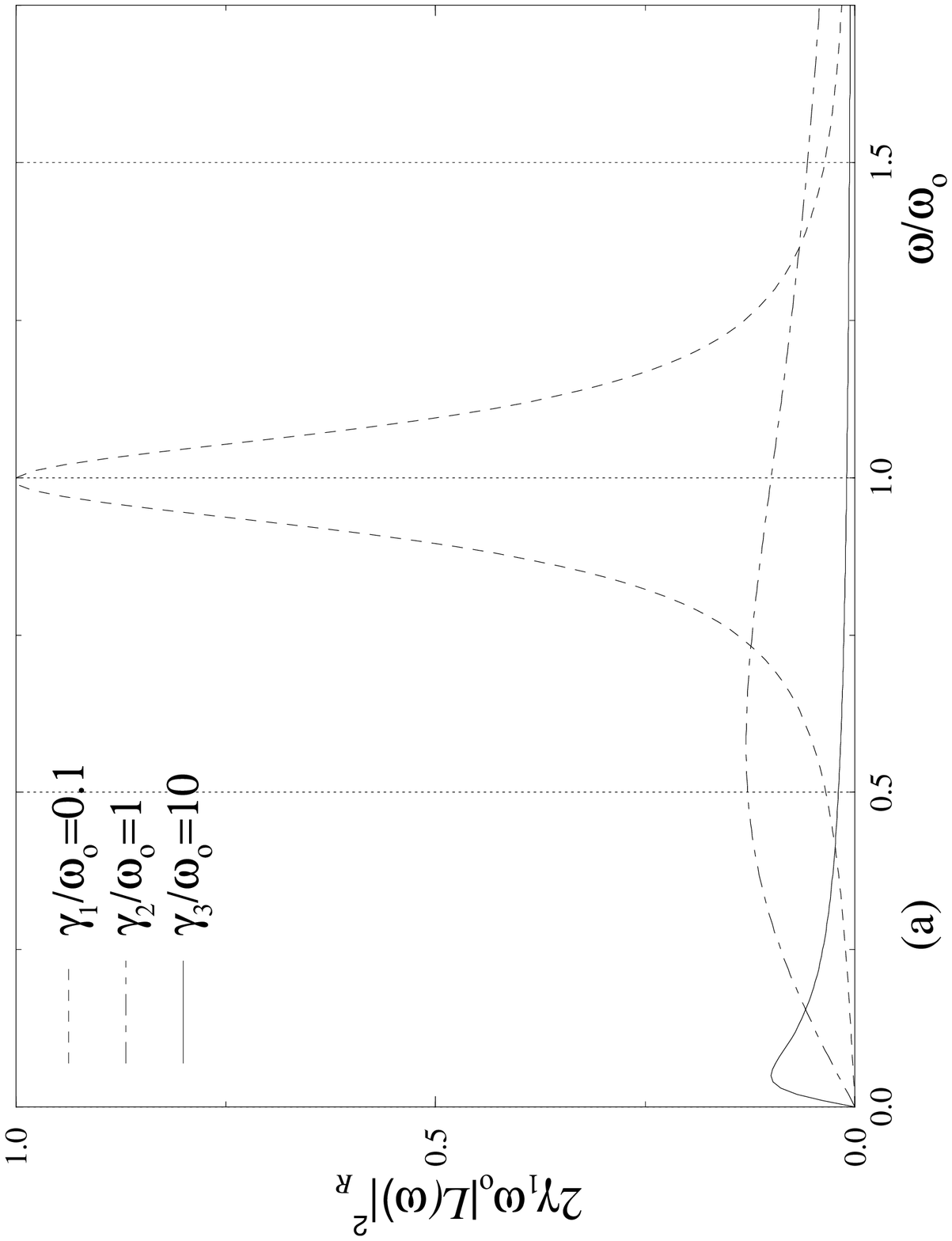}
  }
  \hss}
 }
\end{figure}
\begin{figure}
\vbox to 5.6cm {\vss\hbox to 8.2cm
 {\hss\
   {\includegraphics{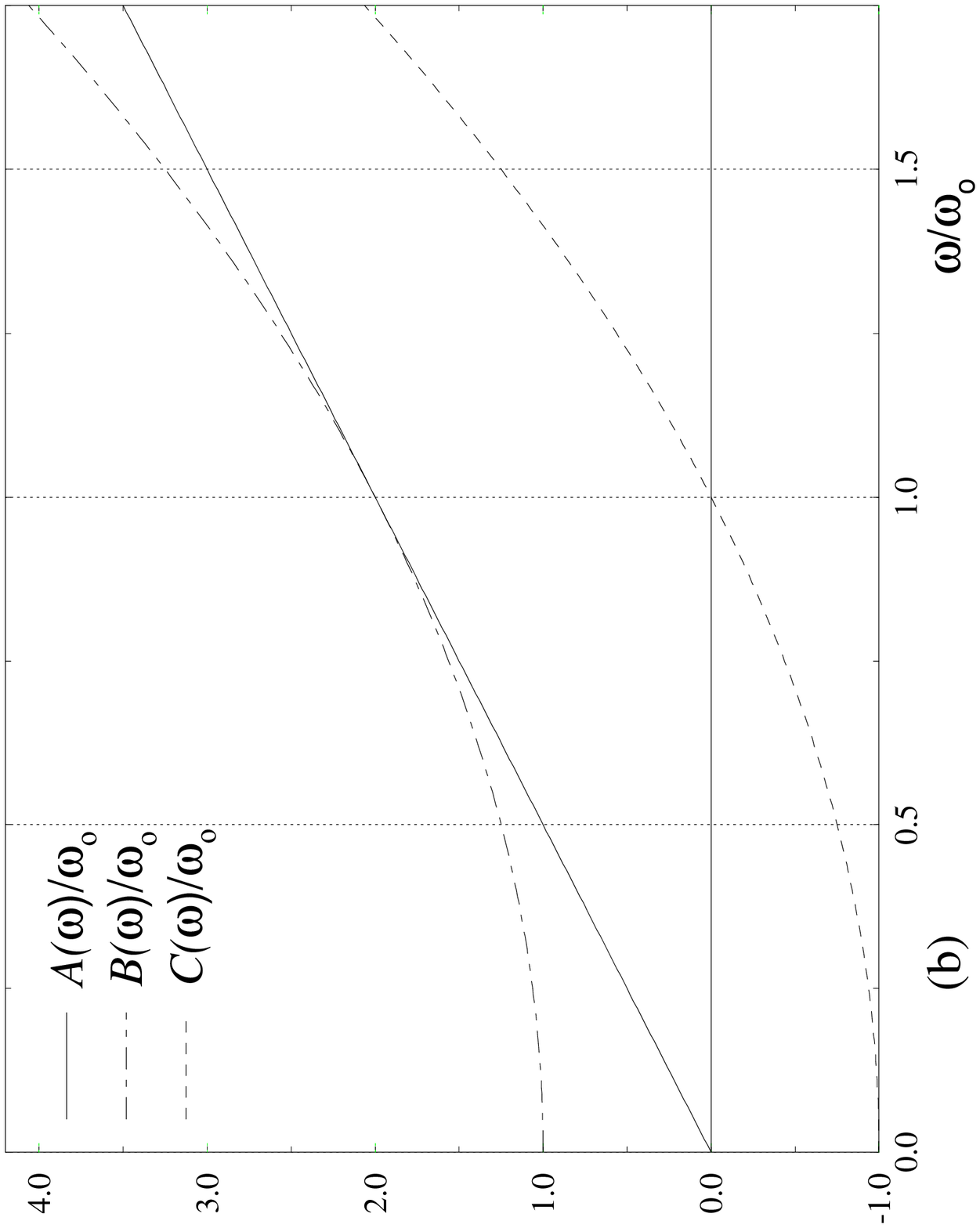}
  }
  \hss}
 }
\caption{(a) Graph of $\left| L\left( \omega \right) \right| _{R}^{2}$ for different rations $%
\gamma /\omega _{o}$. (b)\ Graph of the functions $A\left( \omega \right) $, 
$B\left( \omega \right) $ and $C\left( \omega \right) $ that appear
multiplying $\left| L\left( \omega \right) \right| _{R}^{2}$ in the different terms of
the expression for $\hat{a}\left( t\right) $.}
\label{abs}
\end{figure}
\end{document}